\documentclass[11pt,letterpaper,preprintnumbers,nofootinbib,superscriptaddress,aps,prd]{revtex4-2}

\usepackage[T1]{fontenc}
\usepackage[utf8]{inputenc}
\usepackage{amsmath,amssymb,amsfonts,mathtools,bm}
\usepackage[colorlinks=true, allcolors=blue]{hyperref}
\usepackage[final]{graphicx}
\usepackage{xcolor}
\usepackage{booktabs}
\usepackage{relsize}
\usepackage{lineno}
\usepackage{comment}
\usepackage{braket}
\usepackage{dsfont}
\usepackage{soul}
\usepackage{siunitx} 
\usepackage[caption=false]{subfig}
\usepackage{cancel}

\newcommand{\dd}{\mathrm{d}}
\newcommand{\ketbra}[2]{| #1 \rangle \langle #2 |}

\def\Babar{{\mbox{\slshape B\kern-0.1em{\smaller A}\kern-0.1em B\kern-0.1em{\smaller A\kern-0.2em R}}}}

\setstcolor{red}
\begin{document}

\preprint{TUM-EFT 217/26}

\title{Heavy quarkonium dynamics beyond the binding energy over temperature expansion}

\author{Nora Brambilla}
\email{nora.brambilla@tum.de}
\affiliation{Technical University of Munich, TUM School of Natural Sciences, Physics Department, James-Franck-Strasse 1, 85748 Garching, Germany}
\affiliation{Institute for Advanced Study, Technische Universität München, Lichtenbergstrasse 2 a, 85748 Garching, Germany}
\affiliation{Munich Data Science Institute, Technische Universität München, Walther-von-Dyck-Strasse 10, 85748 Garching, Germany}

\author{Arthur Yi-Ting Lin}
\email{arthur72@mit.edu}
\affiliation{Massachusetts Institute of Technology,
77 Massachusetts Avenue, Cambridge, MA 02139, USA}

\author{Tom Magorsch}
\email{tom.magorsch@tum.de}
\affiliation{Technical University of Munich, TUM School of Natural Sciences, Physics Department, James-Franck-Strasse 1, 85748 Garching, Germany}

\author{Antonio Vairo}
\email{antonio.vairo@tum.de}
\affiliation{Technical University of Munich, TUM School of Natural Sciences, Physics Department, James-Franck-Strasse 1, 85748 Garching, Germany}

\begin{abstract}
We revisit the pNRQCD open-quantum-system description of bottomonium in a quark–gluon plasma, writing the full master equation in a spherical basis and establishing a framework for its numerical treatment. The equation does not rely on an expansion in the ratio of the binding energy over the temperature, while retaining the Born, Markov, and dilute approximations. It depends on suitable convolutions of Coulombic Green's functions with 
real-time chromoelectric correlators, rather than 
a finite set of transport coefficients. It contains jump-operator contributions that enter with both positive and negative signs, so it is not of a Lindblad type. 
We utilize optimal operator rotations to minimize the negative sector and define a positive-sector approximation 
that retains only the optimized positive terms, obtaining in this way a Lindblad equation that can be simulated efficiently with the open-source code QTraj for arbitrary chromoelectric spectral functions. We treat the discarded sector as a systematic uncertainty associated with positivity-breaking effects of the pseudo-Lindblad generator. Positivity breaking effects can be attributed to the approximations made in the derivation of the quarkonium master equation, such as the Born--Markov approximations. Using a spectral model constrained by known infrared and ultraviolet limits, we compute the in-medium width and Bjorken-evolved survival probabilities of the $\Upsilon(1\mathrm{S},2\mathrm{S},3\mathrm{S})$. 
The computations show the convergence of the binding-energy-over-temperature expansion at high temperatures and provide the first large-scale quantum-trajectory simulation results for bottomonium in the low temperature regime. We can thus follow the bottomonium evolution over the full temperature range relevant in heavy-ion collisions.
\end{abstract}

\maketitle

\section{Introduction}
Quantum chromodynamics (QCD) becomes deconfined at high temperatures and high energy densities, where quarks and gluons form the quark-gluon plasma (QGP) --- a strongly coupled state of hot QCD matter.
The QGP is the central object of study in heavy-ion physics. A key signature to access the QGP in heavy-ion collision experiments is quarkonium suppression. 
The phenomenon of quarkonium suppression was proposed by Matsui and Satz~\cite{Matsui:1986dk}, who predicted that a thermal screening of the quark-antiquark potential would lead to a sequential melting of charmonium states with increasing temperature. 
Since then, a large body of research has investigated the suppression of quarkonium states due to the effects of a hot QGP~\cite{Andronic:2024oxz}. In particular, in recent years, an emerging approach, combining non-relativistic effective field theories~\cite{Brambilla:2004jw} with the open quantum system framework~\cite{Breuer:2002pc},  gained traction~\cite{Akamatsu:2020ypb,Yao:2021lus}. 
Exploiting the hierarchy of scales characterizing 
heavy quarkonium, it is possible to derive a quantum master equation for the real-time evolution of the quark-antiquark pair~\cite{Brambilla:2016wgg,Brambilla:2017zei,Blaizot:2017ypk,Sharma:2019xum,Yao:2018nmy,Miura:2019ssi,Akamatsu:2014qsa}. 
This approach differs from the static picture introduced by Matsui and Satz, as it describes the full non-equilibrium evolution of the quarkonium in the medium. 
Moreover, in contrast to semiclassical evolution equations, the open quantum system approach preserves the quantum nature of quarkonium states.

In this work, we focus on bottomonium suppression, where the large bottom-quark mass leads to a separation of scales that makes the application of non-relativistic effective field theories particularly well-suited. 
Bottomonium suppression has been previously studied in the open quantum system framework using potential non-relativistic QCD (pNRQCD). 
In Refs.~\cite{Brambilla:2016wgg,Brambilla:2017zei}, the authors derive a quantum master equation for the non-equilibrium evolution of bottomonium in the QGP. 
Initially, the derived master equation is a Redfield equation, which generally is not of the Lindblad type. 
The authors of Refs.~\cite{Brambilla:2016wgg,Brambilla:2017zei} then use an expansion in the binding energy of the quarkonium over the temperature of the QGP, to which we refer as the $E/(\pi T)$ expansion, to simplify the master equation. 
At leading order (LO) in the $E/(\pi T)$ expansion, which corresponds to the strict Brownian limit, they show that the master equation becomes of the Lindblad type. 
In this case, the evolution depends on few transport coefficients, which are defined in terms of chromoelectric correlators and capture the non-perturbative, strongly-coupled behavior of the medium. 
In Ref.~\cite{Brambilla:2022ynh}, this approach is extended to next-to-leading order (NLO) in $E/(\pi T)$. 
Beyond NLO it is not possible to reconstruct a Lindblad equation. The simulation code QTraj~\cite{Omar:2021kra}, based on the Monte Carlo wave function algorithm~\cite{PhysRevLett.68.580}, enables large-scale simulations of the LO and NLO Lindblad equations and thereby provides phenomenological predictions of quarkonium suppression observables, measured by CMS~\cite{CMS:2018zza,CMS:2023lfu}, ATLAS~\cite{ATLAS:2022exb}, ALICE~\cite{ALICE:2020wwx}, and STAR~\cite{STAR:2022rpk}. 

While the $E/(\pi T)$ expansion enables simulations at high temperatures, we need to go beyond that expansion to describe the bottomonium evolution in the medium at lower temperatures. This is relevant since, as 
the QGP gradually cools down, the bottomonium spends a considerable amount of time at low temperatures. 
In Ref.~\cite{BR:2025lhx}, the non-Markovian evolution was studied by solving the unexpanded Redfield equation. 
However, that work neglects quantum jumps and uses simulations on small lattices. 
Very recently, the authors of Ref.~\cite{Hammou:2026yqh} obtained a universal Lindblad equation that likewise avoids imposing a hierarchy between the temperature and quarkonium energy gaps within the Born--Markov approximation. 
There, a complete numerical solution of the coupled 
color singlet-color octet quark-antiquark pair evolution was left for future work.
In this work, we go beyond those treatments by 
\emph{(i)} including quantum-jump terms, \emph{(ii)} using for the chromoelectric correlator a model consistent with known constraints, \emph{(iii)} carrying out simulations on larger lattices, and \emph{(iv)} studying excited states.

Motivated by the need of an accurate low-temperature description for phenomenological applications, in this work, we study the pNRQCD quantum master equation without expanding in the binding energy over temperature. 
We start from the pNRQCD master equation, which is a Redfield equation, and develop a QTraj-compatible large-scale quantum-trajectory treatment within an optimized positive-sector approximation. This means that, 
based on work from the open quantum system community~\cite{Becker_2023,Becker:2023qvq}, 
we exploit the freedom to choose a jump operator basis 
that minimizes the norm of the negative contributions in the master equation. We then showcase large-scale simulations of the underlying dynamics and compare them to the LO and NLO in $E/(\pi T)$ master equations. 
Our approach relaxes the assumption $\pi T \gg E$, while retaining the Born and Markov approximations. 
By removing the $\pi T \gg E$ requirement, our work thus provides the first large-scale quantum-trajectory simulation of bottomonium dynamics below $190\,$MeV within the pNRQCD master-equation framework.

Using the Born--Markov approximation and after projecting 
the master equation onto spherical harmonics, the time evolution equation for the density matrix $\rho$ of the heavy quark-antiquark pairs in the medium gets the general form 
\begin{equation}
    \frac{\mathrm{d}\rho}{\mathrm{d}t}= -\mathrm{i}[H,\rho]+\sum_{mn}h_{mn}\left(L_n\rho L^\dagger_m-\frac{1}{2}\left\{L^\dagger_mL_n,\rho\right\}\right),
\end{equation}
where $H$ and the jump-operators $L_n$ are operators acting on the heavy quark-antiquark Hilbert space, and the coefficient matrix $h$, which couples the jump-operators, is commonly referred to as the Kossakowski matrix~\cite{Gorini:1975nb}. 
Diagonalizing $h$ yields independent jump-operator contributions weighted by its eigenvalues. 
If some of these eigenvalues are negative, the resulting evolution is not guaranteed to be positive, and the master equation is thus not of the Lindblad form. 
Following Refs.~\cite{Becker:2021xcj,Becker_2023}, we refer to the diagonal representation containing both positive- and negative-weight jump terms as pseudo-Lindblad form.  
Although this master equation is not guaranteed to preserve the positivity of the density matrix throughout the evolution, contributions generating unphysical negative eigenvalues should turn out to be suppressed as long as the Born--Markov approximation remains valid. 
We treat the impact of the negative terms as a systematic uncertainty and study their size in Sec.~\ref{sec:width} and Sec.~\ref{sec:bjorken}. Simulating the full non-Lindblad master equation, including the negative terms, is challenging, as the stochastic unraveling implemented in the QTraj code is not directly applicable. 
While specialized unraveling schemes that solve non-Lindblad master equations have been developed~\cite{Kondov_2003,Breuer:1999ma,Breuer:2004cdm,Donvil:2021bwe,Becker_2023}, computationally, these unraveling algorithms are significantly more demanding than the unraveling algorithms applied to solve Lindblad equations. 
To enable large-scale simulations, we instead apply optimal rotations of the jump operators that minimize the negative terms in the pseudo-Lindblad equation. 
We then define a positive-sector approximation by retaining the optimized positive terms and discarding the residual negative sector. The approximation yields a Lindblad equation that can be simulated efficiently with QTraj. 
This approach for treating negative contributions in a master equation was first introduced in Ref.~\cite{Becker:2021xcj} and realized on an extended Hubbard chain. 
Here, we apply it to the non-equilibrium evolution of bottomonium and demonstrate that the approach is computationally feasible for large-scale simulations. 
The optimize-and-truncate approach is similar in spirit to the approach taken for the NLO in $E/(\pi T)$ equation. 
The difference is that here the hierarchy between the positive and negative sectors is not obtained from the $E/(\pi T)$ power counting, but from an explicit numerical minimization of the negative sector. 
The optimize-and-truncate approach allows us to investigate the dynamics of bottomonium at lower temperatures and compare the results with the expanded LO and NLO master equations. 

Without the $E/(\pi T)$ expansion, the medium effects cannot be absorbed into transport coefficients, but instead depend on suitable convolutions of Coulombic Green's functions with real-time chromoelectric correlators. At leading order in the gauge coupling, 
the chromoelectric correlators are equal to the correlator of two chromoelectric fields connected by an adjoint Wilson line that we call, for short, adjoint correlator~\cite{Nijs:2023dbc,Brambilla:2025xnw}. 
The different correlators were also identified with the adjoint correlator in the LO and NLO formulation.
In the general formulation developed here, we keep the chromoelectric correlators distinct and thus derive the All-Orders master equation without this approximation. 
For the numerical study, however, we adopt the equal correlator approximation to enable a direct comparison with previous work. Perturbatively, the adjoint correlator 
has been computed in real time~\cite{Binder:2021otw,Scheihing-Hitschfeld:2023tuz}, and imaginary time~\cite{Brambilla:2025xnw}. 
While first lattice computations of the Euclidean chromoelectric correlator have been performed~\cite{Datta:2025tpf,Brambilla:2025cqy}, the extraction of the corresponding real-time correlator is hindered by the inverse problem. 
A complementary strategy is to compute the correlator(s) directly in real time, avoiding analytic continuation altogether: lattice simulations have been used in this way to extract gauge-field spectral functions~\cite{Boguslavski:2018beu}, albeit within the classical-statistical approximation and predominantly in highly-occupied, far-from-equilibrium settings, so that a full non-perturbative real-time determination of the thermal adjoint correlator remains an open problem.
In this work, we model the spectral function imposing infrared and ultraviolet constraints, and use the resulting real-time correlator as input to the master equation. In this way, we provide a practical numerical recipe for implementing arbitrary spectral functions in the QTraj simulation by precomputing the necessary operators and using optimal operator rotations.

The paper is organized as follows: 
In Sec.~\ref{sec:quarkoqs}, we revisit the open-quantum-system formulation of bottomonium in medium. 
In Sec.~\ref{sec:master_eq_sph}, we show the projection of the full master equation onto  spherical harmonics and give the final form of the projected master equation in Sec.~\ref{sec:AOeq}. 
To address the negative terms present in this equation, we discuss the optimal operator rotations in Sec.~\ref{sec:optimaloperators}. 
For comparison, we summarize the LO and NLO master equations in the $E/(\pi T)$ expansion in Sec.~\ref{sec:NLO}. 
We demonstrate the application of the master equation in Sec.~\ref{sec:results} by introducing a model for the chromoelectric correlator in Sec.~\ref{sec:correlator} and calculating the in-medium width in Sec.~\ref{sec:width}. 
In Sec.~\ref{sec:bjorken}, we study the evolution of S-wave quarkonium survival probabilities in a Bjorken medium and compare with the results from the LO and NLO master equations derived in Refs.~\cite{Brambilla:2022ynh, Brambilla:2017zei}. 
Finally, in Sec.~\ref{sec:conclusions}, we draw some conclusions.

\section{Theoretical framework and master equation}
\label{sec:theory}
In this part of the paper, we start in Sec.~\ref{sec:quarkoqs} by briefly reviewing the evolution equation previously derived from pNRQCD. To solve the resulting equation not relying upon a specific hierarchy between the binding energy and the temperature, in Sec.~\ref{sec:master_eq_sph}, we project the equation onto spherical harmonics without 
expanding in $E/(\pi T)$. In Sec.~\ref{sec:AOeq}, we then diagonalize the Kossakowski matrix and write the master equation in the corresponding jump-operator representation, which we refer to as pseudo-Lindblad form. 
The final master equation contains jump-operator terms entering with negative signs, which makes the master equation not of the Lindblad form. In Sec.~\ref{sec:optimaloperators}, we discuss how to address these negative terms using operator rotations that minimize their magnitudes. Finally, in Sec.~\ref{sec:NLO}, we summarize the LO and NLO master equations obtained in the $E/(\pi T)$ expansion.

\subsection{Quarkonium as an open quantum system}
\label{sec:quarkoqs}
Heavy quarkonium bound states can be treated nonrelativistically.
This is a consequence of the heavy-quark mass $M$ 
being the largest energy scale in the system,  
which, in turn, implies that the relative velocity of the quark-antiquark pair is small, $v\ll 1$. 
In particular bottomonium exhibits a small relative velocity and, as a consequence, 
a strong separation of nonrelativistic energy scales. The strong separation of scales makes bottomonium well suited to be studies within nonrelativistic effective field theories (EFTs). Potential non-relativistic QCD is obtained from QCD by successively integrating out the hard scale $M$ and the soft scale $Mv$~\cite{Brambilla:1999xf,Brambilla:2004jw,Brambilla:2008cx}. 
In weakly coupled pNRQCD, the soft scale is perturbative, i.e. 
$Mv \gg \Lambda_\text{QCD}$, leading to an EFT with color-singlet and color-octet quark-antiquark fields coupled to ultrasoft gluons as degrees of freedom. 
Ultrasoft gluons have energy and momentum of order $Mv^2$, which is the energy scale of the binding energy $E$.
Due to its small radius $r \sim 1/(Mv)$, bottomonium is an ideal probe of the medium. The radius is small with respect to the inverse of the temperature scale $\pi T$ reached in current heavy-ion collision experiments, so that $\pi T \ll Mv$.
From this it follows that the leading interaction with the medium is encoded into a chromoelectric dipole interaction and the medium dependence factorizes with respect to the bottomonium one. Furthermore, we may assume, in full generality, that the medium is a strongly coupled plasma characterized by no parametric hierarchy among the medium scales, $gT\sim \pi T \sim \Lambda_\text{QCD}$.

In the open quantum system framework, the relevant energy scales define timescales associated with the dynamics of the system, the environment, and their interaction. 
In particular, the intrinsic timescale of the system is given by the binding energy of the bound state, $\tau_S\sim 1/E$, while the timescale of the medium is set by the decay time of the chromoelectric correlator, following from interactions induced by the chromoelectrc dipole term, and is parametrically $\tau_E\sim 1/(\pi T)$. 
A third relaxation timescale $\tau_R$ is generated by the interaction between the system and the medium. 
The Born approximation treats the system–medium coupling perturbatively. The leading order system-medium interaction is given by the chromoelectric dipole interaction.
Higher order terms are suppressed by powers of $g(r \, \pi T)$ and $g (\pi T/M)$. The validity of this approximation is ensured by the small quarkonium size $r \sim 1/(Mv)$ and by the (low) temperature reached at current heavy-ion colliders, so it is fulfilled even when the medium coupling $g$ is large. 
The medium itself is kept nonperturbative, entering through the full chromoelectric correlator, while changes in the medium state induced by the quarkonium are neglected. 
The Markov approximation assumes that the timescale of the medium is much shorter than the relaxation timescale, $\tau_E\ll\tau_R$, so that the medium retains no memory and the evolution becomes local in time.

Using pNRQCD to exploit the hierarchy of energy scales, the authors of Refs.~\cite{Brambilla:2016wgg,Brambilla:2017zei} derived a master equation for the time evolution of the heavy quark-antiquark pair density matrix in a QGP. Heavy quark-antiquark pairs may be in a color singlet or color octet configuration. The corresponding density matrices are denoted $\rho_s$ and $\rho_o$, respectively. Within the Born--Markov approximation and assuming a dilute quarkonium system so that terms nonlinear in the density matrices are neglected, the evolution equations for $\rho_s$ and $\rho_o$ take the form
\begin{align}
	\frac{ \dd\rho_{s}(t)}{ \dd t} &= -i[h_{s}, \rho_{s}(t)] - \Sigma_{s} \rho_{s}(t) - \rho_{s}(t) \Sigma^{\dagger}_{s} + \Xi_{so}(\rho_{o}(t)),\label{eq:general_singlet}\\
	\frac{ \dd\rho_{o}(t)}{ \dd t} &= -i[h_{o}, \rho_{o}(t)] - \Sigma_{o} \rho_{o}(t) - \rho_{o}(t) \Sigma_{o}^{\dagger}
	+ \Xi_{os}(\rho_{s}(t)) + \Xi_{oo}(\rho_{o}(t)),
    \label{eq:general_octet}
\end{align} 
where 
\begin{align}
h_{s(o)} = \frac{p^2}{M} + V_{s(o)}(r) 
\end{align}
is the in-vacuum Hamiltonian of the singlet (octet) with $V_{s(o)}(r)$ the static quark-antiquark color singlet (octet) potential. At leading order, in the adopted hierarchy of energy scales, the static quark-antiquark color singlet and octet potentials are given by the attractive and repulsive Coulomb potentials 
\begin{align}
V_s(r) = - \frac{C_F  \alpha_s}{r} , \qquad \qquad \qquad V_o (r) = \frac{\alpha_s}{2 N_c r}, 
\end{align}
where $N_c=3$ is the number of colors and $C_F = (N_c^2 -1)/(2N_c) = 4/3$. 
The other terms in \eqref{eq:general_singlet} and \eqref{eq:general_octet} are 
\begin{align}
	\Sigma_{s} &= r_{i} A_{i}^{so \dagger}, \label{eq:sigmas}\\
	\Sigma_{o} &= \frac{1}{N_{c}^{2}-1} r_{i} A_{i}^{os \dagger}
	+ \frac{N_{c}^{2}-4}{2(N_{c}^{2}-1)}r_{i}A_{i}^{oo \dagger}, \\
	\Xi_{so}(\rho_{o}(t)) &= \frac{1}{N_{c}^{2}-1} \left( A_{i}^{os \dagger} \rho_{o}(t) r_{i} + r_{i} \rho_{o}(t) A_{i}^{os}\right), \\
	\Xi_{os}(\rho_{s}(t)) &= A_{i}^{so \dagger} \rho_{s}(t) r_{i} 
	+ r_{i} \rho_{s}(t) A_{i}^{so}, \\
	\Xi_{oo}(\rho_{o}(t)) &= \frac{N_{c}^{2}-4}{2(N_{c}^{2}-1)}\left(
	A_{i}^{oo \dagger} \rho_{o}(t) r_{i} + r_{i} \rho_{o}(t) A_{i}^{oo}\right),
    \label{eq:Xi}
\end{align}
where $\bm{r}$ is the position operator, and the subscript $i$ 
identifies the Cartesian coordinate. The operator $A^{uv}_i$ is defined as the convolution integral 
\begin{equation}
    A^{uv}_i = \frac{g^2}{6N_c}\int^\infty _0  \dd s\,\mathrm{e}^{-\mathrm{i}h_u s}r_i\mathrm{e}^{\mathrm{i}h_vs}\mathcal{E}_{uv}(s),
     \label{eq:Aop}
\end{equation}
with $u,v \in \{s,o\}$. The quantities $\mathcal{E}_{uv}(s)$ are gauge-invariant correlators of two chromoelectric fields located at the times $0$ and $s$.
The correlators associated with the different color 
combinations $so$, $os$ and $oo$ differ in the arrangement of the Wilson lines connecting the chromoelectric fields~\cite{Scheihing-Hitschfeld:2022xqx,Brambilla:2025xnw,Brambilla:2024191}. 
The evolution equations~\eqref{eq:general_singlet} and \eqref{eq:general_octet} are master equations that encode the full time evolution of the singlet and octet density matrices. 
The two equations can be written in a joint form in terms of jump operators $L^n_i$ as
\begin{equation}
    \frac{\mathrm{d}\rho(t)}{\mathrm{d}t}
= -\mathrm{i}[H,\rho(t)]
\,+\,\sum_{nm} h_{nm}\!
\left(
  L_i^{n}\rho(t)\,L_i^{m\dagger}
  - \frac{1}{2}\big\{L_i^{m\dagger}L_i^{n},\rho(t)\big\}
\right),
\label{eq:pseudoLindblad}
\end{equation}
where all operators are given in a $2\times 2$ block matrix form, representing the singlet and octet sectors, 
\begin{align}
\rho(t) =
\begin{pmatrix}
\rho_{s}(t) & 0 \\
0 & \rho_{o}(t)
\end{pmatrix},
\qquad\qquad 
&H =
\begin{pmatrix}
h_{s} + \frac{1}{2\mathrm{i}}\big(\Sigma_{s}-\Sigma_{s}^\dag\big) & 0 \\
0 & h_{o} + \frac{1}{2\mathrm{i}}\big(\Sigma_{o}-\Sigma_{o}^\dag\big)
\end{pmatrix}, \label{eq: block form}\\
L_i^{0} =
\begin{pmatrix}
0 & 0 \\
0 & 1
\end{pmatrix} r_i,
\qquad\qquad
&L_i^{1} =
\begin{pmatrix}
0 & 0 \\
0 & \dfrac{N_c^2 - 4}{2\,(N_c^2 - 1)}\,A_i^{oo\dagger}
\end{pmatrix}, \\
L_i^{2} =
\begin{pmatrix}
0 & 1 \\
1 & 0
\end{pmatrix} r_i,
\qquad\qquad
&L_i^{3} =
\begin{pmatrix}
0 & \dfrac{1}{N_c^2 - 1}\,A_i^{os\dagger} \\
A_i^{so\dagger} & 0
\end{pmatrix}.
\end{align}
The Kossakowski matrix $h_{mn}$ mixes the different jump operators $L^n_i$ and is given by 
\begin{equation}
    h =
\begin{pmatrix}
0 & 1 & 0 & 0 \\
1 & 0 & 0 & 0 \\
0 & 0 & 0 & 1 \\
0 & 0 & 1 & 0
\end{pmatrix}.
\label{eq:hmat}
\end{equation}
To bring Eq.~\eqref{eq:pseudoLindblad} into the typical form of a Lindblad equation~\cite{Lindblad:1975ef,Gorini:1975nb}, we need to diagonalize the matrix $h$ and rotate the operators $L^n_i$ to a basis where the evolution equation only contains terms mixing the same jump operators. 
Notice that the paired operators $(L_i^0, L_i^1)$ and $(L_i^2, L_i^3)$ must be rescaled to the same mass dimension before performing this rotation; 
we implement this explicitly in Sec.~\ref{sec:AOeq}. Diagonalizing $h$, however, leads to two eigenvalues $+1$ and two eigenvalues $-1$. 
The negative eigenvalues of the Kossakowski matrix $h$ correspond to terms that break the complete-positivity property of the induced evolution. 
Therefore, Eq.~\eqref{eq:pseudoLindblad} with $h$ as given in Eq.~\eqref{eq:hmat} is not a Lindblad equation.

\subsection{Master equation in the spherical basis}
\label{sec:master_eq_sph}
To study the master equation without assuming a hierarchy between the quarkonium binding energy $E$ and the QGP temperature $T$, we directly project the evolution equations \eqref{eq:general_singlet} and \eqref{eq:general_octet} onto the spherical basis and do not perform an $E/(\pi T)$ expansion. To simplify the notation, we rewrite Eqs. \eqref{eq:general_singlet} and \eqref{eq:general_octet} in the form
\begin{align}
\frac{ \dd \rho_{u}(t)}{ \dd  t} &= -\mathrm{i}[h_{u}, \rho_{u}(t)] - \Sigma_{u} \rho_{u}(t) - \rho_{u}(t) \Sigma^{\dagger}_{u} + \sum_v \Xi_{uv}(\rho_{v}(t)), \label{eq: master eq u}
\end{align}
where $u,v\in \{ s, o\}$ are the color labels. The individual terms are defined in Eqs.~\eqref{eq:sigmas}--\eqref{eq:Xi}, while $\Xi_{ss} = 0$, as there are no singlet-to-singlet transitions mediated by a gluon. 
Anticipating that we shall recast Eq.~\eqref{eq: master eq u} into a pseudo-Lindblad form, we rearrange it as 
\begin{align}
\frac{ \dd \rho_{u}(t)}{ \dd  t} 
& = -\mathrm{i}\left[ h_u + \frac{1}{2\mathrm{i}} \left(\Sigma_u - \Sigma_u^\dag \right), 
\rho_u(t)
\right] - \frac{1}{2} \left\{\Sigma_u + \Sigma_u^\dag , \rho_u \right\} + \sum_v \Xi_{uv} (\rho_v(t)).  \label{eq: master eq rearranged}
\end{align}
The quantity in the commutator can be interpreted as the vacuum Hamiltonian plus a Lamb-shift type correction, 
the anti-commutator is the decay term, and $\Xi_{uv} (\rho_v)$ is the jump term that describes quantum jumps associated to the color transition $v\to u$. 

Since every term in Eq.~\eqref{eq: master eq u} is invariant under rotations, an initially rotationally invariant density matrix remains block diagonal in the angular momentum quantum number $l$ and degenerate in the magnetic quantum number $m$. 
Therefore, it suffices to describe $\rho_u$ using the radial coordinate $r \in \mathbb{R}_{\geq 0}$, the angular momentum quantum number $l \in \mathbb{Z}_{\geq 0}$, and the color $ u$. 
Here, we sketch the projection procedure, which we 
discuss with more details in Appendix~\ref{app:angularprojection}. 

As $\rho_u$ is degenerate in $m$, we can write it as
\begin{align}
\rho_u = \sum_{l=0}^\infty \sum_{m=-l}^{l} \frac{1}{2l+1} \ketbra{lm}{lm} \rho^l_u, 
\end{align} 
where \begin{align}
\rho^l_u = \sum_{m} \bra{lm}\rho_u \ket{lm} =  (2l+1) \bra{lm}\rho_u \ket{lm}, \label{eq:rholu}
\end{align}
with $m$ arbitrarily fixed on the right hand side of Eq.~\eqref{eq:rholu}.
The evolution of each diagonal block $\rho^l_u$ is thus given by 
\begin{align}
 \frac{\dd \rho^l_u}{\dd t} = \sum_m \bra{lm} \frac{\dd \rho_u}{\dd t} \ket{lm}.  
\end{align}
Projecting the terms in Eq.~\eqref{eq: master eq u} onto blocks with angular momentum $l$, we obtain
\begin{align}
\sum_m \bra{lm}\Sigma_u \rho_u \ket{lm} & =  \sum_{v}  c_{u\to v}  \left(
\frac{l+1}{2l+1} r {A^{uv \dag }}_{l+1, l} + \frac{l}{2l+1} r {A^{uv \dag }}_{l-1, l}  \right) \rho^l_u ,  \\
\sum_m \bra{lm}\rho_u \Sigma_u^\dag  \ket{lm} & =  \sum_{v}  c_{u\to v}  \rho^l_u \left(
\frac{l+1}{2l+1}  {A^{uv}}_{l, l+1} r  + \frac{l}{2l+1}  {A^{uv}}_{l, l-1} r  \right)  ,  \\
\begin{split}
\sum_m \bra{lm}\Xi_{uv}(\rho_{v}(t)) \ket{lm} 
& = c_{v\to u} \Bigg[  \frac{l+1}{2l+3} \left({A^{vu\dag}}_{l,l+1} \rho_v^{l+1} r + r \rho_v^{l+1}{A^{vu}}_{l+1,l} \right)   \\
& \qquad  \qquad \quad + \frac{l}{2l-1} \left({A^{vu\dag}}_{l,l-1} \rho_v^{l-1} r + r \rho_v^{l-1}{A^{vu}}_{l-1,l} \right)  \Bigg], 
\end{split}
\end{align}
where 
\begin{align}
c_{s\to s} =0,\qquad\qquad  c_{s\to o} = 1, \qquad\qquad c_{o\to s} = \frac{1}{N_c^2 -1} , \qquad\qquad c_{o\to o} = \frac{N_c^2 -4}{2(N_c^2 -1)}, 
\end{align}
and 
\begin{align}
{A^{uv}}_{l'l} & = \frac{g^2}{6N_c}\int^\infty _0  \mathrm{d}s\,\mathrm{e}^{-\mathrm{i}h^{l'}_u s}r \mathrm{e}^{\mathrm{i}h^{l}_vs}\mathcal{E}_{uv}(s) , \quad {A^{uv \dag }}_{l'l}  = ({A^{uv}}_{ll'})^\dag ,  \label{eq:A}
\end{align}
with the vacuum Hamiltonians in the spherical basis given by 
\begin{equation}
    h^{l}_{u} = -\frac{1}{M}\left(\frac{\partial^2}{\partial r^2} + \frac{2}{r}\frac{\partial}{\partial r}\right) + \frac{l(l+1)}{Mr^2} + V_{u}(r).
\end{equation} 
We then define the jump operators 
\begin{align}
L^{\uparrow/\downarrow}_{u\to v, 0} & =   \mathds{1}_{u\to v} \otimes   O^{\uparrow/\downarrow}   \, \sqrt{N^{\uparrow/\downarrow}} \, r ,   \\
L^{\uparrow/\downarrow}_{u\to v, 1}  &=   \mathds{1}_{u\to v}  \otimes  O^{\uparrow/\downarrow} \, \sqrt{N^{\uparrow/\downarrow}} \, c_{u \to v}  \, {{{A^{uv}}^{\dag}}_{\uparrow/\downarrow},}
\end{align} 
where $\mathds{1}_{u\to v}$ are the $2\times 2$ matrices 
\begin{align}
\mathds{1}_{s \to s} = \begin{pmatrix}
1 & 0 \\
0 & 0 
\end{pmatrix}, \quad \mathds{1}_{s \to o} = \begin{pmatrix}
0 & 0 \\
1  & 0 
\end{pmatrix}, \quad \mathds{1}_{o \to s} = \begin{pmatrix}
0 & 1 \\
0  & 0 
\end{pmatrix}, \quad \mathds{1}_{o \to o} = \begin{pmatrix}
0 & 0 \\
0  & 1 
\end{pmatrix} \label{eq:colortransmatrix}, 
\end{align}
responsible for the transitions from the color state $u$ to the color state $v$, and $O^{\uparrow/\downarrow}$ are the infinite-dimensional matrices $O^{\uparrow/\downarrow}_{l',l} = \delta_{l',l\pm 1}$ that are responsible for increasing/decreasing the angular momentum $l$ by one unit.
The angular momentum factors associated with the two angular momentum transitions are $N^{\uparrow} = (l+1)/(2l+1)$ and $N^{\downarrow} = l/(2l+1)$. 
For the jump operator $L^{\uparrow/\downarrow}_{u\to v, 1}$, there are the additional factors  ${{A^{uv}}^{\dag}}_{\uparrow} = {{A^{uv}}^{\dag}}_{l+1,l}$ and ${{A^{uv}}^{\dag}}_{\downarrow} = {{A^{uv}}^{\dag}}_{l - 1,l}$.
In terms of the jump operators, the terms appearing in Eqs.~\eqref{eq: master eq u} and~\eqref{eq: master eq rearranged} can be written as 
\begin{align}
 \Sigma_u  & = \sum_v \sum_{t = \uparrow, \downarrow} {L^{t}_{u\to v, 0}}^\dag  L^{t}_{u\to v, 1} , \label{eq:lambshiftterm}\\
\Sigma_u^\dag  & = \sum_v \sum_{t = \uparrow, \downarrow} {L^{t}_{u\to v, 1}}^\dag  L^{t}_{u\to v, 0},  \\
\Xi_{uv}(\rho_{v}(t))  & = \sum_{t = \uparrow,\downarrow}  \left( L^{t}_{v\to u, 1}  \rho_v {L^{t}_{v\to u, 0}}^\dag+  L^{t}_{v\to u, 0}  \rho_v {L^{t}_{v\to u, 1}}^\dag\right), \label{eq:jumpterm} 
\end{align} 
which, after substitution into Eq.~\eqref{eq: master eq rearranged}, yields a master equation that can be brought into pseudo-Lindblad form.

After the projection to the spherical basis, each $l$ block has matrix elements that are fully described by the radial coordinates, as $\displaystyle \rho^l(r',r) = \sum_m \bra{r'; lm} \rho \ket{r; lm}$. 
The radial basis vector $\ket{r}$ is normalized with respect to the volume measure $r^2\mathrm{d}r$ and satisfies the orthogonality and completeness conditions  
\begin{align}
\braket{r'|r} = \frac{1}{(r')^2} \delta(r'-r) \quad , \qquad 
\int \dd r \, r^2 \ketbra{r}{r} = 1.   
\end{align} 
Following Ref.~\cite{Brambilla:2022ynh}, we choose the new radial basis vectors $\ket{\bar{r}} = r \ket{r}$, so that it is normalized with respect to the flat radial measure $\mathrm{d}r$ and satisfies
\begin{align}
\braket{\bar{r}'|\bar{r}} = \delta(r'-r) \quad , \qquad 
\int \dd r \,  \ketbra{\bar{r}}{\bar{r}} = 1.   
\end{align}
The matrix elements of the density matrix in terms of the new radial basis read
\begin{align}
\rho^l(\bar{r}', \bar{r}) = \sum_m \bra{\bar{r}'; lm}   \rho \ket{\bar{r}; lm}  = r'  r \rho^{l} (r',r). 
\end{align}
We call this new radial basis the reduced spherical representation. In the reduced spherical representation, the radial trace of operators simplifies to 
\begin{align}
\text{Tr}_r (\rho\, \mathcal{O}) = \int \dd r\,r^2 \,  \bra{r} \rho \, \mathcal{O}  \ket{r} = \int \dd r \bra{\bar{r}} \rho \, \mathcal{O}  \ket{\bar{r}}. 
\end{align}
In addition, differential operators are modified according to
\begin{equation}
\int \mathrm{d} r \, r^2 \ket{r} \frac{\partial}{\partial r} \bra{r} =  \int \mathrm{d} r \ket{\bar{r}} \left(\frac{\partial}{\partial r} - \frac{1}{r} \right) \bra{\bar{r}} , 
\end{equation}
and 
\begin{equation}
\int \mathrm{d} r \, r^2  \ket{r} \left( - \frac{\partial^2}{\partial r^2} - \frac{2}{r} \frac{\partial}{\partial r}\right) \bra{r} = \int \mathrm{d} r \ket{\bar{r}} \left( - \frac{\partial^2}{\partial r^2} \right) \bra{\bar{r}} .
\end{equation}
The last equality is relevant for the kinetic energy operator and implies that the vacuum Hamiltonians $h^l_u$ are written in the reduced spherical representation as 
\begin{align}
h^l_u & = - \frac{1}{M} \frac{\partial^2}{\partial r^2} + \frac{l(l+1)}{Mr^2} + V_u(r). 
\end{align}  
In the rest of the paper, we work in the reduced spherical representation.

After projecting onto the spherical basis and transforming to the reduced radial representation, the jump operator terms can be arranged into a $2\times 2$ color-matrix block form, where the upper diagonal block corresponds to the singlet state and the lower diagonal block corresponds to the octet state.
Following Eqs.~\eqref{eq:lambshiftterm}-\eqref{eq:jumpterm}, we rearrange the master equation~\eqref{eq: master eq rearranged} in an off-diagonal Lindblad-like equation
\begin{align}
\frac{\dd \rho }{\dd t} = -\mathrm{i}\big[H ,\rho \big] + \sum_{u,v = s,o} \sum_{t = \uparrow, \downarrow} \sum_{\alpha, \beta \in \{0,1\}} \sigma_x^{\alpha \beta}\!
\left(
L_{u\to v, \alpha}^{t}\rho \,{L_{u\to v, \beta}^{t}}^\dagger-\frac{1}{2}\big\{ {L_{u\to v,\beta}^{t}}^\dagger L_{u\to v,\alpha}^{t },\rho \big\}
\right) , \label{eq:off_diagonal_Lindblad}
\end{align}
where the Hamiltonian is given by
\begin{align}
H & = \begin{pmatrix}
H_s & 0 \\
0 & H_o
\end{pmatrix}
\end{align}
and $\sigma_x^{00}=\sigma_x^{11}=0$, $\sigma_x^{01}=\sigma_x^{10}=1$. The color-matrix blocks $H_s$ and $H_o$ are diagonal in each angular-momentum block, with components in the reduced spherical representation
\begin{align}
H^l_u = h^l_u + \frac{1}{2\mathrm{i}}\sum_v c_{u\to v} \left(
N^\uparrow \left( r {{A^{uv}}^\dag}_{l+1,l} - {A^{uv}}_{l,l+1} r \right)  + N^\downarrow \left( r {{A^{uv}}^\dag}_{l-1,l} - {A^{uv}}_{l,l-1} r) \right)
\right).
\label{eq: Hamiltonian}
\end{align} 
Using Eq.~\eqref{eq:off_diagonal_Lindblad}, the width operator corresponding to a given color transition $u\to v$ and angular momentum transition $\uparrow/\downarrow$ reads
\begin{equation}
\Gamma^{\uparrow/\downarrow}_{u\to v}  
= \sum_{\alpha,\beta \in \{0,1\}}\sigma_x^{\alpha \beta}  \, {L_{u\to v, \alpha}^{ \uparrow/\downarrow}}^\dag L_{u\to v,\beta}^{\uparrow/\downarrow}  \\
=  c_{u \to v} N^{\uparrow/\downarrow} \left(r {{A^{uv}}^{\dag}}_{l\pm 1,l} + 
{A^{uv}}_{l,l\pm 1} r 
\right). 
\label{eq: all order width}
\end{equation}
In Appendix~\ref{app:NLO expansion}, we show how to recover the NLO width operators obtained in Ref.~\cite{Brambilla:2022ynh}, up to an $\mathcal{O}(E^2/(\pi T)^2)$ difference.

\subsection{All-Orders master equation}
\label{sec:AOeq}
To separate the positive and negative jump sectors of the full master equation, we perform an operator rotation that diagonalizes the Kossakowski matrix $\sigma_x$. 
Before doing so, we notice that the two operators coupled by $\sigma_x$ in Eq.~\eqref{eq:off_diagonal_Lindblad} have different mass dimensions: $L^t_{u\to v,0}\propto r$ has mass dimension $-1$, while $L^t_{u\to v,1}\propto {A^{uv}}^\dagger$ has mass dimension $2$. 
To perform the operator rotation, we therefore first define two jump operators that have the same mass dimension by introducing an arbitrary energy scale $\mu_J$:
\begin{equation}
    \tilde{L}^t_{u\to v,0} \equiv  {\mu_J}^{3/2}L^t_{u\to v,0}, \qquad\qquad\qquad \tilde{L}^t_{u\to v,1} \equiv {\mu_J}^{-3/2}L^t_{u\to v,1}. \label{eq:colortrans}
\end{equation}
Both rescaled operators have mass dimension $1/2$. 
Since $\sigma_x$ is off-diagonal, the factors of $\mu_J$ cancel in every term of Eq.~\eqref{eq:off_diagonal_Lindblad}, so that this rescaling leaves the master equation unchanged. The arbitrary scale $\mu_J$ reflects the non-uniqueness of the jump-operator decomposition for a given master equation. We then diagonalize $\sigma_x$ in the rescaled operator basis, which brings the master equation~\eqref{eq:off_diagonal_Lindblad} for the evolution of the quarkonium density matrix in the reduced spherical representation into pseudo-Lindblad form
\begin{equation}
    \frac{\mathrm{d}\rho(t)}{\mathrm{d}t}
= -\mathrm{i}[H,\rho(t)]
+ \sum_{u,v=s,o}\sum_{t=\uparrow,\downarrow}\sum_{\sigma = \pm}\sigma\left(
{\mathcal{C}}^{t,\sigma}_{u\to v}\rho(t){{\mathcal{C}}^{t,\sigma}_{u\to v}}^\dagger
  - \frac{1}{2}\bigl\{{{\mathcal{C}}^{t,\sigma}_{u\to v}}^\dagger {\mathcal{C}}^{t,\sigma}_{u\to v}, \rho(t)\bigr\}
\right).
\label{eq:finalAO}
\end{equation}
The Hamiltonian $H$ in the reduced spherical basis is block diagonal in angular momentum and color, with blocks given by
\begin{align}
    {H}^l = \begin{pmatrix}
        {h}^l_s + \frac{1}{2\mathrm{i}}\left(\Sigma^l_s - {\Sigma^l_s}^\dag\right)& 0 \\ 0 & {h}^l_o +  \frac{1}{2\mathrm{i}}\left(\Sigma^l_o - {\Sigma^l_o}^\dag\right)
        \end{pmatrix}, 
\label{eq:hamiltonian}
\end{align}
where 
\begin{align}
{h}^l_{s,o}& =-\frac{1}{M}\frac{\partial^2}{\partial r^2}+\frac{l(l+1)}{Mr^2}+V_{s,o}(r),\\
\Sigma^l_s &= \frac{l+1}{2l+1} r {A^{so}}^\dagger_{l+1,l} + \frac{l} {2l+1} r {A^{so}}^\dagger_{l-1,l}, \\
\begin{split}
\Sigma^l_o 
& = \frac{l+1}{2l+1}  \left(
\frac{1}{N_c^2 -1} r {A^{os}}^\dagger_{l+1,l} + \frac{N_c^2 -4}{2(N_c^2 -1)}
r {A^{oo}}^\dagger_{l+1,l} 
\right) \\ &+ \frac{l}{2l+1}  \left(
\frac{1}{N_c^2 -1} r {A^{os}}^\dagger_{l-1,l} + \frac{N_c^2 -4}{2(N_c^2 -1)}
r {A^{oo}}^\dagger_{l-1,l}
\right).
\end{split}
\end{align} 
In the following, we refer to the master equation \eqref{eq:finalAO} as the \textit{All-Orders (AO) master equation}, where All-Orders means
not having expanded in $E/(\pi T)$. 
The jump operators ${\mathcal{C}}^{{\uparrow/\downarrow},\sigma}_{u\to v}$ in Eq.~\eqref{eq:finalAO} are defined as
\begin{equation} {\mathcal{C}}^{{\uparrow/\downarrow},\sigma}_{u\to v} = \mathds{1}_{u\to v}\otimes O^{{\uparrow/\downarrow}} \,  {C}^{{\uparrow/\downarrow}\sigma}_{u\to v},
\end{equation}
with $\mathds{1}_{u\to v}$ the $2\times 2$ matrices for the respective color transitions listed in Eq.~\eqref{eq:colortransmatrix}, the matrices $O^{\uparrow/\downarrow}$ defined after Eq.~\eqref{eq:colortransmatrix}, and the twelve individual jump components
given by
\begin{align}
    &{C}^{\uparrow \pm}_{s\to o}=\frac{1}{\sqrt{2}}\sqrt{\frac{l+1}{2l+1}}\left( \mu^{3/2}_Jr\pm \mu^{-3/2}_J{A^{so}}^\dagger_{l+1,l}\right),\label{eq:jump1}\\
    &{C}^{\downarrow \pm}_{s\to o}=\frac{1}{\sqrt{2}}\sqrt{\frac{l}{2l+1}}\left( \mu^{3/2}_Jr\pm \mu^{-3/2}_J{A^{so}}^\dagger_{l-1,l}\right),\label{eq:jump2}\\
    &{C}^{\uparrow \pm}_{o\to s}=\frac{1}{\sqrt{2}}\sqrt{\frac{l+1}{2l+1}}\left( \mu^{3/2}_Jr\pm\frac{1}{N^2_c-1}\mu^{-3/2}_J{A^{os}}^\dagger_{l+1,l}\right),\label{eq:jump3}\\
    &{C}^{\downarrow \pm}_{o\to s}=\frac{1}{\sqrt{2}}\sqrt{\frac{l}{2l+1}}\left( \mu^{3/2}_Jr\pm\frac{1}{N^2_c-1} \mu^{-3/2}_J {A^{os}}^\dagger_{l-1,l}\right),\label{eq:jump4}\\
    &{C}^{\uparrow \pm}_{o\to o}=\frac{1}{\sqrt{2}}\sqrt{\frac{l+1}{2l+1}}\left( \mu^{3/2}_Jr\pm\frac{N^2_c-4}{2(N^2_c-1)}\mu^{-3/2}_J{A^{oo}}^\dagger_{l+1,l}\right),\label{eq:jump5}\\
    &{C}^{\downarrow \pm}_{o\to o}=\frac{1}{\sqrt{2}}\sqrt{\frac{l}{2l+1}}\left( \mu^{3/2}_Jr\pm\frac{N^2_c-4}{2(N^2_c-1)}\mu^{-3/2}_J{A^{oo}}^\dagger_{l-1,l}\right),
    \label{eq:jump6}
\end{align} 
where $\mu_J$ is the arbitrary energy scale introduced above. From the jump operators, we define the signed contribution of each jump operator to the width as
\begin{align}
    \Gamma^{{\uparrow/\downarrow},\sigma}_{u\to v} \equiv \sigma {{\mathcal{C}}^{{\uparrow/\downarrow},\sigma}_{u\to v}}^\dagger {\mathcal{C}}^{{\uparrow/\downarrow},\sigma}_{u\to v}.
    \label{eq:signedwidth}
\end{align}
The full width for each color transition $u\to v$ is then given by the sum
\begin{equation}
    \Gamma_{u\to v}=\sum_{t=\uparrow, \downarrow
    }\sum_{\sigma=\pm }   \Gamma^{t,\sigma}_{u\to v} .
    \label{eq:widthCdagC}
\end{equation}
The individual positive- ($\sigma = +1$) and negative-sector ($\sigma = -1$) contributions in Eq.~\eqref{eq:signedwidth} depend on the choice of $\mu_J$, whereas their sum in Eq.~\eqref{eq:widthCdagC} does not. 
Indeed, the terms proportional to $\mu^3_Jr^2$ and $\mu^{-3}_J {A^{uv}}{A^{uv}}^\dagger$ cancel between the two sectors, leaving only cross terms, which are independent of $\mu_J$. Carrying out the summation in Eq.~\eqref{eq:widthCdagC} explicitly, we obtain the expressions of the widths for the three possible color transitions 
\begin{align}
\Gamma_{s\to o} & = \frac{l+1}{2l+1} \left(r {A^{so}}^\dagger_{l+1,l} + {A^{so}}_{l,l+1}r \right) + \frac{l}{2l+1} \left(r {A^{so}}^\dagger_{l-1,l} + {A^{so}}_{l,l-1} r \right), \label{eq:singletwidth}
 \\
\Gamma_{o\to s} & =  \frac{1}{N_c^2-1} \left[\frac{l+1}{2l+1} \left(r {A^{os}}^\dagger_{l+1,l} + {A^{os}}_{l,l+1} r \right) + \frac{l}{2l+1} \left(r {A^{os}}^\dagger_{l-1,l} + {A^{os}}_{l,l-1} r \right) 
\right], \label{eq:octetsingletwidth} \\
\Gamma_{o\to o} &= \frac{N_c^2-4}{2(N_c^2-1)} \left[\frac{l+1}{2l+1} \left(r {A^{oo}}^\dagger_{l+1,l} + {A^{oo}}_{l,l+1} r \right) + \frac{l}{2l+1} \left(r {A^{oo}}^\dagger_{l-1,l} + {A^{oo}}_{l,l-1} r \right) 
\right].\label{eq:octetoctetwidth}
\end{align} 
As expected, the arbitrary scale $\mu_J$ cancels from Eqs.~\eqref{eq:singletwidth}-\eqref{eq:octetoctetwidth}. 
Thus, not only the full AO master equation but also its full width operator is independent of the normalization scale $\mu_J$. Expanding the exponentials that appear in we recover the decay width from the NLO master equation  of Ref.~\cite{Brambilla:2022ynh} up to positivity-restoring terms of $\mathcal{O}(E^2/(\pi T)^2)$; we show this in Appendix~\ref{app:NLO expansion}.

\subsection{Optimal operator rotations}
\label{sec:optimaloperators}
The All-Orders master equation~\eqref{eq:finalAO} contains jump terms with a negative sign and is therefore not of Lindblad form. Equivalently, the time-local AO generator is not guaranteed to generate a completely positive dynamical map for arbitrary initial states and evolution times. 
Because the positivity of the density matrix is not guaranteed to be preserved, the master equation can lead to negative eigenvalues in the density matrix, which are unphysical. 
These unphysical negative eigenvalues are, however, expected to be small if the approximations made in the derivation of the master equation are sufficiently well satisfied. 
In our case, these approximations are first and foremost the Born and Markov approximations, which are expected to hold sufficiently well due to the separation of scales in the system. 
While the negative terms corresponding to $\sigma=-1$ in Eq.~\eqref{eq:finalAO} can generate such negative eigenvalues in the density matrix throughout the evolution, they do not necessarily do so for a given initial state and timeframe. 
They can thus contain both physical and unphysical contributions. 
We propose treating the size and impact of the negative sector as a systematic uncertainty, quantified by the difference between the full master equation and its positivity-restoring truncation. The uncertainty could arise from the approximations made when deriving the full master equation, such as the Born--Markov approximation. 
We investigate the size of the negative contributions in Sec.~\ref{sec:width} and Sec.~\ref{sec:bjorken}.

In principle, it is possible to unravel master equations of the form~\eqref{eq:finalAO} including the negative terms. However, because the negative terms can yield negative probabilities in the naive quantum unraveling framework, specialized unraveling schemes are required. 
Non-positive master equations can be unraveled by enlarging the Hilbert space, in particular using doubled-Hilbert-space constructions~\cite{Kondov_2003,Breuer:1999ma} and tripled-space embeddings~\cite{Breuer:2004cdm}. Recently, novel unraveling schemes based on weighted trajectory averages with signed contributions have been proposed~\cite{Donvil:2021bwe,Becker_2023}. 
While these unraveling schemes enable the simulation of the full master equation, including the negative terms, they all require significant computational resources, making large-scale simulations challenging. We further comment on the implementation of the unraveling scheme~\cite{Becker_2023} for the simulation of the All-Orders master equation in Appendix~\ref{app:plqt}. 

Here, we focus on a slightly different approach. As discussed in Ref.~\cite{Brambilla:2022ynh} and the following Sec.~\ref{sec:NLO}, at high temperatures, a suitable rotation of the jump operators leads to negative terms that are suppressed in $E/(\pi T)$ and can be therefore neglected in the NLO master equation. 
Similarly, we expect the negative contributions to be small at lower temperatures as well, however, we cannot rely on the $E/(\pi T)$ expansion to show this. Instead, we use the idea of optimal operator rotations introduced in Refs.~\cite{Becker_2023,Becker:2023qvq}. 
Originally, these operator rotations were used to mitigate the sign problem induced by the weighted average in the pseudo-Lindblad Quantum Trajectories unraveling scheme of Ref.~\cite{Becker_2023} and discussed in Appendix~\ref{app:plqt}. 
The idea is that any master equation of the form 
\begin{equation}
    \frac{\dd\rho(t)}{dt} = -\mathrm{i}[H,\rho(t)] + \sum_{\sigma = \pm} \sigma \left(L_\sigma \rho(t) L^\dagger_\sigma -\frac{1}{2}\left\{L^\dagger_\sigma L_\sigma,\rho\right\}\right)
\end{equation}
is invariant and remains in pseudo-Lindblad form under transformations $W$ that rotate the jump operators as
\begin{equation}
    \begin{pmatrix}
        L^\prime_+ \\ L^\prime_-
    \end{pmatrix} = W 
    \begin{pmatrix}
        L_+ \\
        L_-
    \end{pmatrix}, 
    \label{eq:operatorrotation}
\end{equation}
if conjugation by $W$ preserves the diagonalized coefficient matrix $J=\mathrm{diag}(1,-1)$, that is 
\begin{align}
W^\dag J W = J . 
\end{align}
We call this class of transformations pseudo-unitary.\footnote{The transformations would be unitary if $J$ would be proportional to the identity.}
It was shown in \cite{Becker:2023qvq} that, up to independent overall phases of the transformed jump operators, the most general form of $W$, is given by 
\begin{equation}
    W = \begin{pmatrix}
        e^{\mathrm{i}\phi/2}\cosh(w) & e^{-\mathrm{i}\phi/2}\sinh(w)\\
        e^{\mathrm{i}\phi/2}\sinh(w) & e^{-\mathrm{i}\phi/2}\cosh(w)
    \end{pmatrix}, \label{eq:Wmatrix}
\end{equation}
with $w$ and $\phi$ two real parameters. 
While the master equation is formally invariant under this rotation, the individual positive- and negative-sign contributions are not. 
This freedom can be used to choose a representation in which the negative sector is minimized. 
If the minimized negative sector is sufficiently small, one can construct a Lindblad approximation from the optimized representation. 
In this approximation, the positive sector is retained, while the residual negative contribution is discarded, in analogy to the NLO master equation. We aim to find the operator rotation that minimizes the negative terms, thereby making this positive-sector approximation as accurate as possible. 
Furthermore, we expect that an optimal operator rotation leads to smaller negative terms and, therefore, to a more accurate approximation even at high temperatures, where the $E/(\pi T)$ expansion holds.

To find an optimal operator rotation, we 
minimize the squared norm of the negative jump operator after rotation $||L^\prime_-||^2$. 
In principle, different choices for the norm are possible. 
Here, we consider the choice 
\begin{equation}
    ||L^\prime_-||^2 = \text{Tr}\left[\rho_\text{w}{L^\prime_-}^\dagger L^\prime_-\right],
    \label{eq:norm}
\end{equation}
where $\rho_\text{w}$ is a weight operator, which can be chosen according to the relevant region of Hilbert space that we aim to describe. 
For $\rho_\text{w}=\mathds{1}$, this expression becomes the squared Frobenius norm, while other choices of $\rho_\text{w}$ allow to assign a stronger weight to different parts of the Hilbert space. 
We notice that for $\rho_\text{w}>0$ Eq.~\eqref{eq:norm} is the square of a norm, while for $\rho_\text{w}\geq0$ it is only a semi-norm, lacking definiteness, 
however this is not relevant for the following discussion. 

Minimizing Eq.~\eqref{eq:norm} with respect to $\phi$ and $w$, we obtain 
\begin{align}
    \phi &= \pi - \arg\left[\text{Tr}\left(\rho_\text{w}L^\dagger_-L_+\right)\right],\label{eq:phi}\\
    w &= \frac{1}{2}\tanh^{-1}\left(\frac{2\left|\text{Tr}\left(\rho_\text{w}L^\dagger_-L_+\right)\right|}{||L_+||^2+||L_-||^2}\right).\label{eq:w}
\end{align}
Given any pair of jump operators $L_+,L_-$, we can obtain the optimal jump operators $L^\prime_+,L^\prime_-$ from Eq.~\eqref{eq:operatorrotation} with parameters given by \eqref{eq:phi} and \eqref{eq:w}. 
We then expect $||L^\prime_+||^2 \gg ||L^\prime_-||^2$, allowing us to approximate the evolution by only retaining the positive sector of the optimized representation. 
We show that  $||L^\prime_+||^2 \gg ||L^\prime_-||^2$ holds in our simulations of bottomonium evolution in Sec.~\ref{sec:bjorken}. 
The positive-sector approximation leads to a Lindblad equation that can be simulated efficiently using the standard wave-function Monte Carlo algorithm implemented in QTraj~\cite{Omar:2021kra}.
We denote this Lindblad equation as \textit{positive All-Orders (AO$+$) master equation.}

We apply the described operator optimization to all operator pairs $C^{t\pm}_{u\to v}$ in Eqs.~\eqref{eq:jump1}-\eqref{eq:jump6}. 
In particular, we only perform rotations among jump operators that induce the same color and angular-momentum transitions, so that the optimized representation does not mix different transition channels. 
This ensures that each trajectory can still be represented by a single wave function carrying definite angular momentum and color quantum numbers, thereby preserving the favorable computational scaling of the quantum trajectories algorithm. 

The arbitrary normalization scale $\mu_J$ introduced in Eqs.~\eqref{eq:jump1}-\eqref{eq:jump6} does not affect the operator optimization. 
For a fixed color and angular-momentum transition, we put the corresponding positive- and negative-sign operators into a vector
\begin{equation}
    \vec{C}^t_{u\to v}(\mu_J) = \begin{pmatrix}
        C^{t+}_{u\to v}(\mu_J)\\
        C^{t-}_{u\to v}(\mu_J)
    \end{pmatrix}.
\end{equation} 
Considering two different scale choices $\mu_J$ and $\mu^\prime_J$, we define
\begin{equation}
    \eta=\frac{3}{2}\ln\frac{\mu^\prime_J}{\mu_J}.
\end{equation}
Since ${\mu^\prime}^{\pm 3/2}_J=e^{\pm\eta} \mu^{\pm 3/2}_J$, the jump operators $\vec{C}^t_{u\to v}$ at the two different scales are related by
\begin{equation}
    \vec{C}^t_{u\to v}(\mu^\prime_J)=S(\eta)\vec{C}^t_{u\to v}(\mu_J),
\end{equation}
where the transformation $S(\eta)$ can be written as
\begin{equation}
    S(\eta)=\begin{pmatrix}
        \cosh\eta & \sinh\eta \\
        \sinh\eta & \cosh\eta 
    \end{pmatrix}.
\end{equation}
Thus, changing the scale $\mu_J$ is itself a transformation of the form~\eqref{eq:Wmatrix}, corresponding to $w=\eta$ and $\phi=0$,
i.e. a unitary transformation that leaves the pseudo-Lindblad form of the master equation invariant. 

The optimized representation is also independent of $\mu_J$. If $W_{\mu_J}$ denotes an optimal transformation~\eqref{eq:Wmatrix} for the operators $\vec{C}^t_{u\to v}(\mu_J)$ at a scale $\mu_J$, then at a different scale $\mu^\prime_J$ we can choose
\begin{equation}
    W_{\mu^\prime_J} = W_{\mu_J}S(-\eta),
\end{equation}
so that 
\begin{equation}
    W_{\mu^\prime_J}\vec{C}^t_{u\to v}(\mu^\prime_J) = W_{\mu_J}\vec{C}^t_{u\to v}(\mu_J).
\end{equation}
Since both $W_{\mu_J}$ and $S(-\eta)$ belong to the same class of unitary transformations that leave the pseudo-Lindblad form invariant, their product $W_{\mu^\prime_J}=W_{\mu_J}S(-\eta)$ does so as well. 
When expressed in the phase convention used in Eq.~\eqref{eq:Wmatrix}, this product may additionally contain independent overall phases of the final positive- and negative-sector jump operators. 
Such rephasings, $C_\pm^\prime\to e^{i\theta_\pm}C_\pm^\prime$, leave both the corresponding dissipative terms and the norm in Eq.~\eqref{eq:norm} unchanged. 
Therefore, the minimum of the negative-sector squared norm and the optimized jump operators are independent of $\mu_J$, up to physically irrelevant phases. Consequently, the AO$+$ generator obtained by discarding the optimized negative sector is independent of the arbitrary normalization scale $\mu_J$.
The argument holds for any fixed choice of the weight operator $\rho_{\rm w}$. 

As previously discussed, we have the freedom to choose the weight operator $\rho_{\rm w}$.
In the following, we use $\rho_\text{w}=\mathds{1}$ as the main weight operator. 
Since this choice of $\rho_\text{w}$ is positive and the calculation of $||L^\prime_\pm||^2$ is performed on a finite lattice, the operators have positive and finite norms, ensuring that the optimization procedure is well defined. Possible alternatives could consist in weighting the physically most relevant region of the Hilbert space by choosing, e.g., a thermal state $\rho_\text{w}\propto \exp(-H/T)$, or the phenomenologically most important states, e.g., $\rho_\text{w}\propto \ketbra{1\mathrm{S}}{1\mathrm{S}}+\ketbra{2\mathrm{S}}{2\mathrm{S}} + \ketbra{3\mathrm{S}}{3\mathrm{S}}$. 
In general, the choice of $\rho_\text{w}$ can alter the magnitude of the negative contributions for different parts of the Hilbert space. We further discuss the impact of the choice of $\rho_{\rm w}$ in Sec.~\ref{sec:width}.

\subsection{Master equations in the $E/(\pi T)$ expansion}
\label{sec:NLO}

For comparison with the AO equation, we summarize the LO and NLO master equations obtained by expanding in $E/(\pi T)$. 
In Refs.~\cite{Brambilla:2016wgg,Brambilla:2017zei,Brambilla:2022ynh}, the correlators $\mathcal{E}_{uv}(s)$ were approximated by the adjoint chromoelectric correlator describing color singlet-to-octet transitions 
\begin{equation}
    \mathcal{E}_{uv}(s) \approx \mathcal{E}_{so}(s) = \braket{E^a_j(0)\Omega^{ab}(0,s)E^b_j(s)},
\end{equation}
where $\Omega^{ab}(0,s)$ denotes an adjoint Wilson line connecting the two chromoelectric fields. 
At leading order in the gauge coupling, the different correlators coincide, while beyond leading order they differ through their different Wilson-line structures~\cite{Scheihing-Hitschfeld:2022xqx,Brambilla:2025xnw,Brambilla:2024191}.
To simplify the evolution equations further, in Refs.~\cite{Brambilla:2016wgg,Brambilla:2017zei,Brambilla:2022ynh} it was assumed that the correlator $\mathcal{E}_{so}(s)$ decays sufficiently rapidly to vanish for $s\gg 1/(\pi T)$. In this case and assuming the hierarchy $\pi T \gg E$, the exponentials in Eq.~\eqref{eq:Aop} can be expanded as the Hamiltonian is of the order of the binding energy $h_u\sim E$. 
This expansion corresponds to the quantum Brownian regime $\tau_E\ll\tau_S$. 
In Refs.~\cite{Brambilla:2016wgg,Brambilla:2017zei}, a master equation was derived by setting the exponentials to one. 
In Ref.~\cite{Brambilla:2022ynh}, the approach was extended to include next-to-leading order terms in the $E/(\pi T)$ expansion. We refer to these equations as the \textit{LO and NLO master equations}, respectively. In both cases, contributions that arise from the negative eigenvalues of the Kossakowski matrix $h$, which  are $\mathcal{O}\left(E^2/(\pi T)^2\right)$, 
can be neglected as they lie beyond the intended accuracy of the master equation. 
Hence, performing the $E/(\pi T)$ expansion up to NLO yields a genuine Lindblad equation given by
\begin{equation}
    \frac{\mathrm{d}\rho(t)}{\mathrm{d}t}
= -\mathrm{i}[H,\rho(t)]
+ \sum_{n=0}^{1}\left(
  C_i^{n}\rho(t)C_i^{n\dagger}
  - \frac{1}{2}\bigl\{C_i^{n\dagger} C_i^{n}, \rho(t)\bigr\}
\right),
\label{eq:LindNLO}
\end{equation}
with
\begin{equation}
H = \begin{pmatrix}
h_s + \frac{r^{2}}{2}\gamma
+ \frac{\kappa}{4 M T}\{ r_{i}, p_{i} \} & 0 \\
0 & h_o + \frac{N_{c}^{2}-2}{2\bigl(N_{c}^{2}-1\bigr)}
\left(
  \frac{r^{2}}{2}\gamma
  + \frac{\kappa}{4 M T}\{ r_{i}, p_{i} \}
\right)
\end{pmatrix} \label{eq: Hamiltonian NLO}
\end{equation}
and
\begin{align}
\begin{split}
    &C_i^{0} =
\sqrt{\frac{\kappa}{N_c^{2}-1}}\,
\begin{pmatrix}
0 & 1 \\
0 & 0
\end{pmatrix}
\left(
r_i + \frac{\mathrm{i} p_i}{2 M T} + \frac{\Delta V_{os}}{4 T}\, r_i
\right)\\
&\qquad + \sqrt{\kappa}\,
\begin{pmatrix}
0 & 0 \\
1 & 0
\end{pmatrix}
\left(
r_i + \frac{\mathrm{i} p_i}{2 M T} + \frac{\Delta V_{so}}{4 T}\, r_i
\right),
\end{split}\\
&C_i^{1} =
\sqrt{\frac{\kappa \bigl(N_c^{2}-4\bigr)}{2\bigl(N_c^{2}-1\bigr)}}\,
\begin{pmatrix}
0 & 0 \\
0 & 1
\end{pmatrix}
\left(
r_i + \frac{\mathrm{i} p_i}{2 M T}
\right),
\end{align}
where $\Delta V_{uv} \equiv V_u(r) - V_v(r)$. This particularly simple form follows from also assuming the instantanous approximation discussed in Appendix \ref{app:NLO expansion}; 
this approximation is however not necessary for getting a Lindblad equation.
In \eqref{eq:LindNLO}, there are only two collapse operators per dimension, $C_i^0$ and $C_i^1$, which correspond to the positive eigenvalues of $h$, giving a total of six jump operators. 
The operators corresponding to the two negative eigenvalues of $h$ are neglected as they contribute at $\mathcal{O}(E^2/(\pi T)^2)$. In this formulation (including the instantaneous approximation), the integral over the correlator in Eq.~\eqref{eq:Aop} 
reduces to two transport coefficients, $\kappa$ and $\gamma$, which are defined via
\begin{align}
    \frac{g^2}{6N_c}\int^\infty_0\mathrm{d}s\, \braket{E^a_j(s)\Omega^{ab}(s,0)E^b_j(0)}=\frac{\kappa+\mathrm{i}\gamma}{2}.
\end{align}
To perform simulations, the evolution equation~\eqref{eq:LindNLO} is projected onto spherical harmonics, see Ref.~\cite{Brambilla:2022ynh}. 
Assuming isotropy, one obtains the evolution of the density matrix $\rho^l_u(r',r)$, which is block diagonal in the color label $u$ and the angular momentum $l$. 
The jump operators act on the radial part of the Hilbert space and induce transitions between color and angular momentum states. 
In particular, $C^0_i$ is responsible for singlet to octet and octet to singlet transitions, while $C^1_i$ for octet to octet transitions. Each transition can change the angular momentum from $l$ to $l+1$ or $l-1$. 
Since the density matrix is block diagonal in $u$ and $l$, the jump terms decouple, and, after the projection onto spherical harmonics, $C^0_i$ and $C^1_i$ lead to six distinct jump operators in the spherical basis. 
For more details on the NLO master equation, we refer the reader to Ref.~\cite{Brambilla:2022ynh}. 
The solution of the Lindblad equation with these six jump operators was implemented in the open-source code QTraj~\cite{Omar:2021kra}.

\section{Application of the all-orders master equation}
\label{sec:results}
In this section, we study the numerical implementation and application of the All-Orders master equation. 
To evaluate the integral \eqref{eq:A}, which determines the jump operators, we need the real-time chromoelectric correlators beyond their lowest moments. 
To show the feasibility of our numerical framework, and to make direct contact with previous solutions based on the $E/(\pi T)$ expansion in Refs.~\cite{Omar:2021kra,Brambilla:2022ynh,Brambilla:2023hkw}, we use the approximation $\mathcal{E}_{uv}(s)\approx \mathcal{E}_{so}(s)$ introduced above for all transition channels. 
We then construct a physical model for the spectral function of this correlator, constrained by the known infrared and ultraviolet behavior. 
Using this model, we present the in-medium decay width of the $1\mathrm{S}$ bottomonium state, comparing AO, AO$+$, LO, and NLO calculations with lattice-QCD extractions. 
Finally, we simulate the survival probabilities of the $\Upsilon(1\mathrm{S})$, $\Upsilon(2\mathrm{S})$ and $\Upsilon(3\mathrm{S})$ states under a Bjorken temperature evolution. 
We first compare the effective-Hamiltonian evolution, in which the quantum jump terms $C\rho C^\dagger$ are neglected, for the AO, AO$+$, LO, and NLO equations. We then include quantum jumps for the AO$+$ Lindblad evolution and compare it with the corresponding LO and NLO evolutions. 
A phenomenological analysis including hydrodynamics and feeddown is deferred to future work.

\subsection{Real-time chromoelectric correlator}
\label{sec:correlator}
We write the Hermitian conjugate of the singlet-to-octet chromoelectric correlator,  $\mathcal{E}^{{\dagger}}_{so}(s)$, which enters the jump operators through ${A^{uv}}^\dagger$, in terms of the spectral function $\rho(\omega)$ as 
\begin{equation}
    \braket{E^a_j(s)\Omega^{ab}(s,0)E^b_j(0)}=\frac{1}{2\pi}\int^\infty_{-\infty}\dd \omega\,e^{-\mathrm{i}\omega s}\rho(\omega)(1+n_B(\omega)),
    \label{eq:correlator}
\end{equation}
where $n_B(\omega) = 1/(e^{\omega/T}-1)$ is the Bose--Einstein distribution. 
As a minimal ansatz, we restrict $\rho(\omega)$ to its odd component, thereby neglecting asymmetric contributions to the correlator, which have been identified in perturbation theory and recent lattice studies~\cite{Scheihing-Hitschfeld:2023tuz,Brambilla:2025xnw,Brambilla:2025cqy}. 
In perturbation theory, these even contributions to the spectral function only enter at next-to-leading order, while the leading-order spectral function is odd. We note that for an odd spectral function, the transport coefficients $\kappa$ and $\gamma$ can be written in terms of this correlator as 
\begin{align}
    \frac{g^2}{6N_c}&\int^\infty_0\mathrm{d}s\, \braket{E^a_j(s)\Omega^{ab}(s,0)E^b_j(0)}=\frac{\kappa+\mathrm{i}\gamma}{2},\label{eq:cond1}\\
    \mathrm{i}\frac{g^2}{6N_c}&\int^\infty_0\dd s\,s\braket{E^a_j(s)\Omega^{ab}(s,0)E^b_j(0)}=\frac{\kappa}{4T}\label{eq:cond2}+\mathrm{i}\tilde{\gamma}.
\end{align}
Here, the second relation~\eqref{eq:cond2} can have a non-zero imaginary part beyond the instantaneous limit assumed in~\cite{Brambilla:2022ynh} and discussed in Appendix~\ref{app:NLO expansion}, which we denote $\tilde{\gamma}$~\cite{Brambilla:2024191}. 
We generally express the transport coefficients in terms of the dimensionless quantities $\hat\kappa$ and $\hat\gamma$, such that $\kappa \equiv \hat\kappa T^3$ and $\gamma \equiv \hat\gamma T^3$, while $\tilde\gamma \equiv \hat{\tilde\gamma}T^2$.

To obtain the correlator from Eq.~\eqref{eq:correlator}, we need a parametrization of the spectral function $\rho(\omega)$, which satisfies the conditions~\eqref{eq:cond1} and~\eqref{eq:cond2}. 
In general, the spectral function is not known over the full range of $\omega$. 
However, in the infrared (IR) limit, the behavior of the spectral function is fixed by the transport coefficient $\hat\kappa$~\cite{Brambilla:2019tpt}, while in the ultraviolet (UV), it can be obtained perturbatively~\cite{Burnier:2010rp,Brambilla:2026213}. 
We therefore fix
\begin{equation}
    \rho_\text{IR}(\omega) = \frac{6N_c}{g^2}\hat\kappa T^2 \omega,\qquad\qquad 
    \rho_\text{UV}(\omega) = \frac{N^2_c-1}{\pi} \omega^3.
\end{equation}
The behavior of the spectral function in between $\rho_\text{IR}$ and $\rho_\text{UV}$ is not known. 
Here, we pick a simple model to smoothly interpolate between the IR and UV parts and leave a more sophisticated study of the spectral function to future work. In particular, we pick a smooth function $s(\omega) \in C^\infty$, which satisfies 
\begin{equation}
    \lim_{\omega\to 0}s(\omega) = 0, \qquad\qquad 
    \lim_{\omega\to\infty} s(\omega) = 1.
\end{equation}
We then define the full spectral function $\rho(\omega)$ as 
\begin{equation}
    \rho(\omega) = \left(1-s(\omega)\right)\rho_\text{IR}(\omega) + s(\omega)\rho_\text{UV}(\omega),
\end{equation}
so that it satisfies the desired IR and UV limits. 
Different choices of smooth $s(\omega)$ are possible, however, we note that the switch $s(\omega)$ must be even to maintain the imposed oddness of $\rho(\omega)$. Our choice is 
\begin{align}
s(\omega)=\frac{\displaystyle \tanh\left(\frac{\omega^2-\chi^2}{2\Delta\chi}\right) + \tanh\left(\frac{\chi}{2\Delta}\right)}{\displaystyle 1+\tanh\left(\frac{\chi}{2\Delta}\right)},
\label{eq:tanhansatz}
\end{align}
with $\Delta$ parameterizing the width of the transition interval between $\rho_\text{IR}$ and $\rho_\text{UV}$, and $\chi$ giving the position of the transition window. 
The normalization ensures that this choice satisfies the required limits for $s(\omega)$. 
We choose $\chi$ and $\Delta$ proportional to the temperature. 
We briefly discuss the dependence of the result on this choice in Appendix~\ref{app:spec}.

The UV tail of the spectral function, which is proportional to $\omega^3$, is responsible for the vacuum contribution to the chromoelectric correlator, which is proportional to $1/s^4$.
The vacuum part of the chromoelectric correlator leads to divergences in Eqs. \eqref{eq:cond1} and \eqref{eq:cond2} and in the evolution equation that need to be subtracted and renormalized, for instance, by the potentials.
Several subtraction schemes are possible.
Our choice for the present work is to define the subtracted chromoelectric correlator as 
\begin{equation}
\braket{E^a_j(s)\Omega^{ab}(s,0)E^b_j(0)}_\text{th} = \braket{E^a_j(s)\Omega^{ab}(s,0)E^b_j(0)} - \braket{E^a_j(s)\Omega^{ab}(s,0)E^b_j(0)}_\text{UV},
\end{equation}
with 
\begin{equation}
\braket{E^a_j(s)\Omega^{ab}(s,0)E^b_j(0)}_\text{UV}=\frac{1}{2\pi}\int^\infty_{-\infty}\dd \omega\,\mathrm{e}^{-\mathrm{i}\omega s} \, \Theta(\omega) \, s(\omega,T) \, \rho_\text{UV}(\omega) .
\end{equation}
This choice guarantees that the UV part is subtracted only at large $\omega$, however it makes the subtraction scheme temperature dependent.
As the main focus in this study is to show the feasability of the numerical solution of the evolution equation also in the absence of a specific hierarchy between binding energy and temperature, we will not elaborate further on our or other choices of subtraction scheme, which we leave to future work.
Simply, from now on we work with the UV-sutracted correlator given by
 
\begin{equation}
\braket{E^a_j(s)\Omega^{ab}(s,0)E^b_j(0)}_\text{th}=\frac{1}{2\pi}\int^\infty_{-\infty}\dd \omega\,\mathrm{e}^{-\mathrm{i}\omega s} \left[\rho(\omega) - \Theta(\omega) \, s(\omega,T)\,\rho_\text{UV}(\omega)  + \rho(\omega)n_B(\omega)\right]. \label{eq: thermal correlator}
\end{equation}
In our model, the correlator depends on $\hat\kappa$, $\chi$, and $\Delta$, which we fix, while the coefficients $\hat\gamma$ and $\tilde\gamma$ are outputs of the model. 
For a given $\hat\kappa$, the resulting correlator satisfies the real parts of the conditions~\eqref{eq:cond1} and~\eqref{eq:cond2}.

\begin{figure}
    \centering
\includegraphics[width=0.49\linewidth]{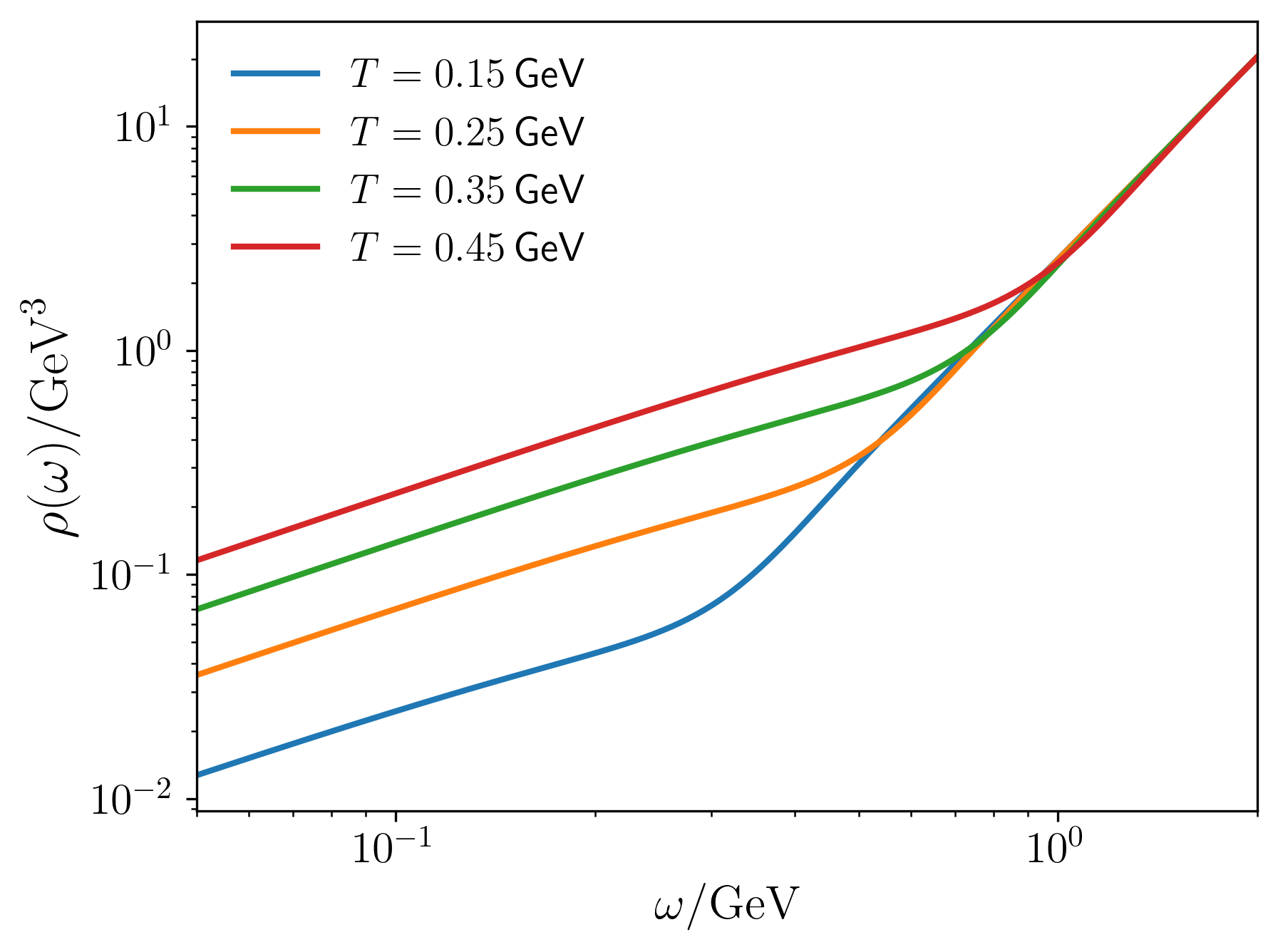}
\includegraphics[width=0.49\linewidth]{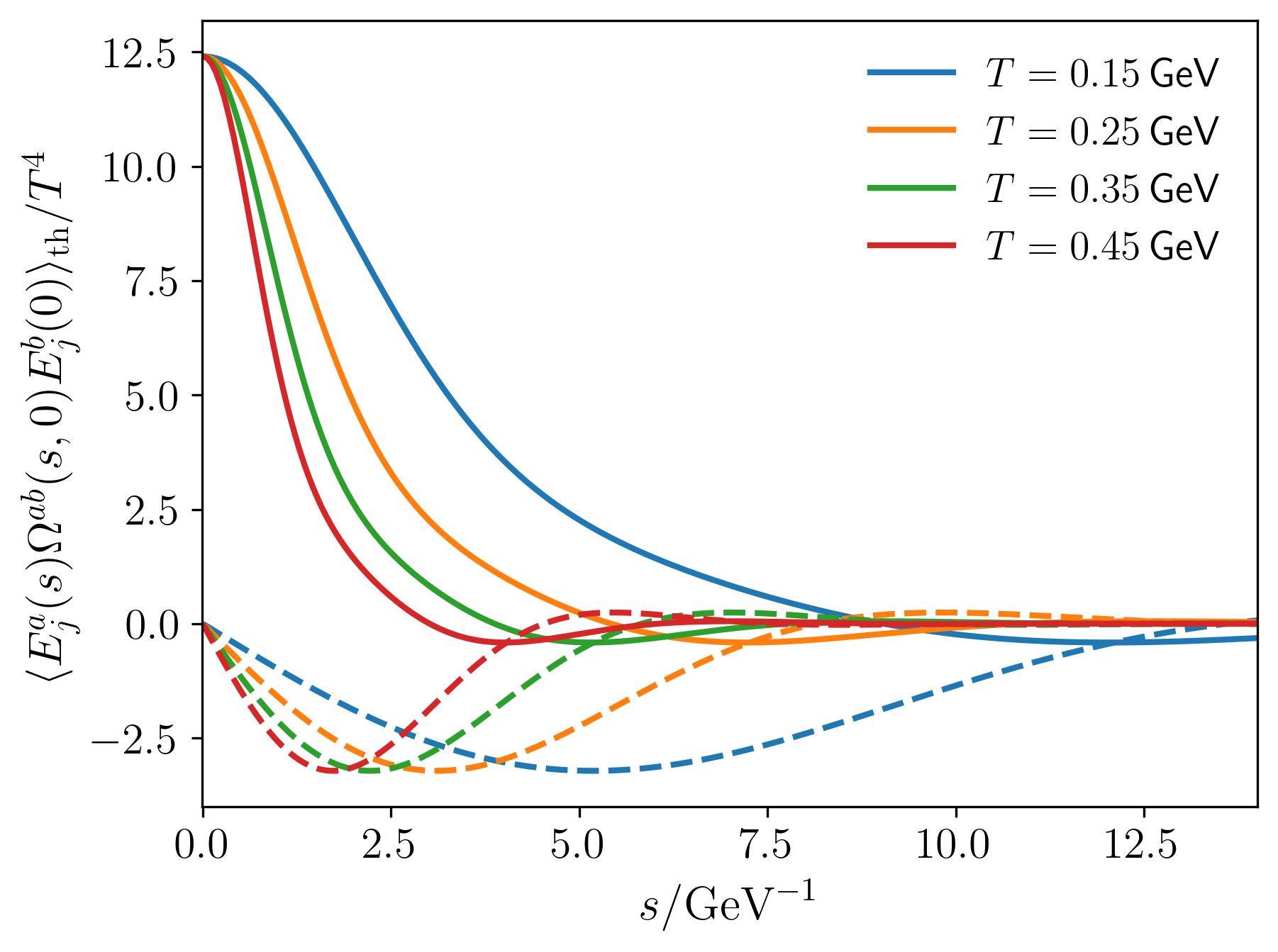}
    \caption{\textit{Left:} The spectral function $\rho(\omega)$ for $\hat\kappa = 2.5$, $\Delta=\chi=(\pi/2)\,T$ and $g^2$ evaluated at the central scale $\mu=1.5\pi T_\text{ref}$ at different temperatures. 
    \textit{Right:} The UV-subtracted correlator divided by the fourth power of the temperature, at different temperatures. 
    The solid lines show the real part of the correlator, while the dashed lines show the imaginary part.}
    \label{fig:rho}
\end{figure}

\begin{figure}
    \centering
\includegraphics[width=0.5\linewidth]{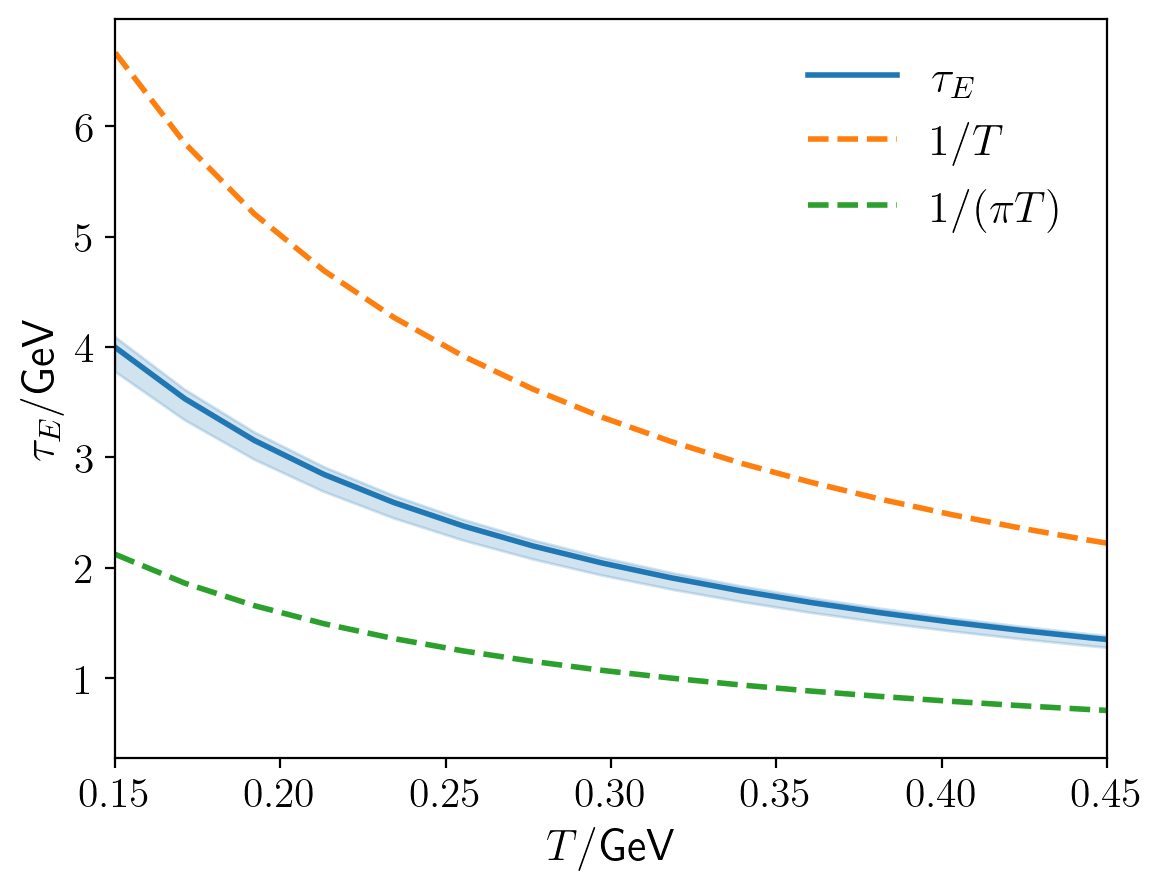}
    \caption{We show the temperature dependence of the chromoelectric memory timescale $\tau_E$ in our model with $\Delta =\chi = (\pi/2)\,T$, and the curves $1/T$ and $1/(\pi T)$. 
    The blue band is obtained by varying the scale $\mu$ used to evaluate the coupling $g^2$ between $\pi T_\text{ref}$ and $2\pi T_\text{ref}$, with the reference temperature chosen as $T_\text{ref}=300\,$MeV.
    The central value corresponds to $1.5\pi T_\text{ref}$.}
    \label{fig:tauE}
\end{figure}

In the following, we pick $\hat\kappa = 2.5$ inspired by the lattice QCD values for the heavy quark diffusion coefficient around $T=400\,$MeV~\cite{HotQCD:2025fbd}. 
We notice that $\hat\kappa$ is expected to depend on the temperature, but for simplicity and in accordance with previous works~\cite{Brambilla:2023hkw}, we keep it constant in this study. 
Since $\hat\gamma$ and $\hat{\tilde{\gamma}}$ are calculated from our spectral function, they can be temperature dependent as a consequence of our ansatz.
Furthermore, we choose $\Delta=\chi=(\pi/2 )\,T$, so that the transition between the IR and UV regimes occurs over a thermal frequency range. 
We discuss the dependence on this choice for $\Delta$ and $\chi$ in Appendix~\ref{app:spec}. 
For the coupling $g^2$ entering the spectral function, we use a fixed-coupling prescription in which $g^2$ is evaluated at a reference scale $\mu\in [\pi T_\text{ref},2\pi T_\text{ref}]$, with $T_\text{ref}=300\,$MeV chosen as a representative temperature scale for the evolution. 
The central value is taken at $\mu=1.5\pi T_\text{ref}$.
We show the resulting spectral function $\rho(\omega)$ for these parameters at different temperatures in the left panel of Fig.~\ref{fig:rho}. 
The corresponding UV-subtracted correlators are shown in the right panel.
We observe that the correlator decays rapidly in time, with higher temperatures leading to faster decays. 
To quantify how quickly the correlator decays, we define the \textit{chromoelectric memory timescale} $\tau_E$ from the absolute first moment of the correlator 
\begin{equation}
    \tau_E = \frac{\displaystyle \int^\infty_0 \dd s\;s\left|\braket{E^a_j(s)\Omega^{ab}(s,0)E^b_j(0)}_\text{th}\right|}{\displaystyle \int^\infty_0 \dd s \left|\braket{E^a_j(s)\Omega^{ab}(s,0)E^b_j(0)}_\text{th}\right|}.
\end{equation}
In the open quantum system framework, we identify $\tau_E$ with the bath correlation time.
We show the dependence of $\tau_E$ on the temperature for our spectral function model in Fig.~\ref{fig:tauE}. 
There, we vary the scale $\mu$ at which we evaluate the coupling between $\pi T_\text{ref}$ and $2\pi T_\text{ref}$ again with $T_\text{ref}=300\,$MeV and show the induced uncertainty as a shaded band. 
The central value corresponds to $1.5\pi T_\text{ref}$.
We observe, as expected, that the timescale approximately decays as $1/(\xi T)$. 
For the factor of the scale, $\xi$, we approximately find $\xi \approx 1.7$. 
This lies below the commonly used thermal estimate $\xi\approx\pi$, which motivates the $E/(\pi T)$ expansion. 
The numerical values of $\tau_E$ are model-dependent, which we illustrate in Appendix~\ref{app:spec}. 
We defer a systematic study of the dependence on the spectral function to future work.

As mentioned previously, while $\hat\kappa$ is a parameter in our model of the spectral function, $\hat\gamma$ and $\hat{\tilde\gamma}$ are outputs that depend on the shape of the spectral function. 
Calculating $\hat\gamma$ from Eq.~\eqref{eq:cond1} for different temperatures, we find that $\hat\gamma(T)\approx-1.55$ with only weak temperature dependence in the range considered. 
For $\hat{\tilde\gamma}$, we find $\hat{\tilde\gamma}\approx 0.33$, again with negligible temperature dependence.  

Our model of the chromoelectric correlator differs in an important way from the simple exponential ansatz used, e.g., in Ref.~\cite{BR:2025lhx}. 
The simple exponential does not, in general, enforce the moment relation in Eq.~\eqref{eq:cond2}, 
even setting $\tilde{\gamma}=0$.
This makes a direct comparison to the NLO-expanded dynamics in the form of Sec.~\ref{sec:NLO} problematic for such an ansatz, since the NLO master equation of Sec.~\ref{sec:NLO} is derived assuming that Eq.~\eqref{eq:cond2} (for $\tilde{\gamma}=0$) holds.

\subsection{In-medium quarkonium width}
\label{sec:width}
\begin{figure}
    \centering
    \includegraphics[width=0.6\linewidth]{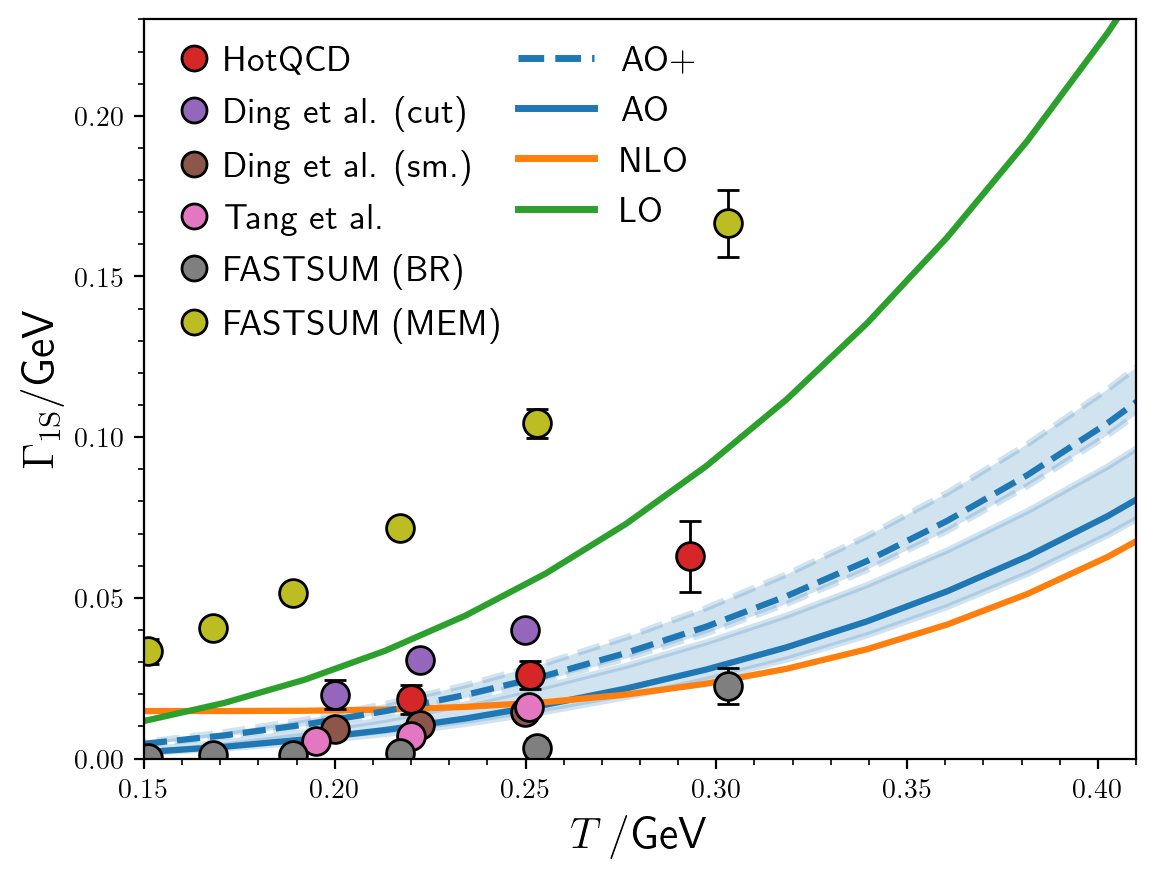}
    \caption{The $1\mathrm{S}$ in-medium width in terms of the temperature for $\hat\kappa = 2.5$. 
    We show the full All-Orders width, including negative contributions, as a blue solid line, and the positive widths obtained from the Frobenius norm as a dashed line. 
    The blue band is obtained by evaluating the coupling $g^2$ at $\pi T_\text{ref}$ and $2\pi T_\text{ref}$, with the reference temperature chosen as $T_\text{ref}=300\,$MeV.
    The central value corresponds to $1.5\pi T_\text{ref}$. 
    Other norm choices do not differ significantly from the Frobenius norm. As a reference, we show the LO and NLO widths. 
    The lattice-QCD points are taken from HotQCD~\cite{Ali:2025iux}, Ding~et~al.~\cite{Ding:2025fvo}, shown for two parametrizations of the in-medium spectral function, a cut Lorentzian (cut) and a smooth-cut Lorentzian (sm.), FASTSUM~\cite{Skullerud:2026sek}, shown for the Bayesian-reconstruction (BR) and maximum-entropy (MEM) methods, and Tang~et~al.~\cite{Tang:2024dkz}.}
    \label{fig:width}
\end{figure}

Using the correlator model introduced in Sec.~\ref{sec:correlator}, we calculate the in-medium width of the $\Upsilon(1\mathrm{S})$ state, given by
\begin{equation}
\Gamma_{1\mathrm{S}}=\braket{1\mathrm{S}|\Gamma_{s\to o}|1\mathrm{S}}, 
\end{equation}
where $\ket{1\mathrm{S}}$ is the ground state of the vacuum singlet Hamiltonian and $\Gamma_{s \to o}$ is the width operator, which we obtain from Eq.~\eqref{eq:singletwidth}. 
In practice, to calculate the width, we discretize the radial space into a finite box of size $L=40\,\rm{GeV}^{-1}$, with $N=1500$ points. 
We checked that, for this choice, finite-lattice-spacing and finite-volume effects on the width are negligible. 
In accordance with previous work, for the bottom-quark mass we choose $M = 4.73\,$GeV fixed from the $\Upsilon(1\mathrm{S})$ mass and set the coupling of the potential to $\alpha_s=0.468$, following the Bohr-radius prescription of Ref.~\cite{Brambilla:2022ynh}. 
More details on the computational setup are given in Refs.~\cite{Omar:2021kra,Brambilla:2022ynh}. 
To evaluate the width operator, we first calculate the operator ${A^{so}}^\dagger_{l+1,l}$ by solving the integral~\eqref{eq:A} numerically at different temperatures, using the UV-subtracted correlator~\eqref{eq: thermal correlator}. 
Then, using Eq.~\eqref{eq:singletwidth}, we calculate $\Gamma_{s\to o}$ at $l=0$ to obtain the width for the full All-Orders master equation. 
In the following, we refer to this width as the \textit{full All-Orders (AO) width}, emphasizing that in this calculation, we also include the negative terms. 
Furthermore, we calculate a width, which we refer to as the \textit{positive All-Orders (AO$+$) width}, by performing the optimal operator rotations and then retaining only the positive sector of the master equation. 
We do this by performing the operator optimization outlined in Sec.~\ref{sec:optimaloperators} on the jump operators $C^{\uparrow\pm}_{s\to o}$ in Eq.~\eqref{eq:jump1} and calculating the width directly according to Eq.~\eqref{eq:widthCdagC}, where we only include positive operators after the optimization. This corresponds to the effective width in the AO$+$ positive-sector approximation. For the numerical evaluation we choose $\mu_J=1\,\mathrm{GeV}$.
As shown in Sec.~\ref{sec:optimaloperators}, the optimized AO$+$ width is independent of this arbitrary normalization scale. 
We calculate this positive All-Orders width for three choices of the weight operator 
\begin{align}
\rho_\text{w}=\mathds{1}, \qquad 
\rho_\text{w} \propto e^{-h_s/T},\qquad 
\rho_\text{w} \propto \ketbra{1\mathrm{S}}{1\mathrm{S}}+\ketbra{2\mathrm{S}}{2\mathrm{S}}+\ketbra{3\mathrm{S}}{3\mathrm{S}}  .
\end{align}
We compare the obtained widths to the ones calculated from the width operators at LO and NLO in the $E/(\pi T)$ expansion, which are given in Appendix~\ref{app:NLO expansion}.
The results are shown in Fig.~\ref{fig:width}, where we compare to extractions from lattice-QCD and lattice-QCD-informed determinations~\cite{Ali:2025iux,Ding:2025fvo,Skullerud:2026sek,Tang:2024dkz}. Since different bottom-quark masses were used in the lattice calculation, we rescale the width data by $(M^\mathrm{lattice}/M)^2$, where $M^\mathrm{lattice}$ denotes the bottom-quark mass in the respective lattice calculation and $M$ denotes the bottom-quark mass used in our calculation. 
We again vary the scale $\mu$ used to fix the coupling $g^2$ between $\pi T_\text{ref}$ and $2\pi T_\text{ref}$ with $T_\text{ref}=300\,$MeV. 
The blue bands show the resulting variation, while the central values correspond to $1.5\pi T_\text{ref}$.
We observe that the difference between the full AO and NLO widths is much smaller than the difference between the NLO and LO widths, consistent with the convergence of the $E/(\pi T)$ expansion at sufficiently high temperatures. 
We note that the All-Orders widths decrease rapidly at low temperatures, whereas the LO and NLO widths are notably larger.\footnote{ 
What we call NLO in Fig.~\ref{fig:width} includes NNLO terms in the $E/(\pi T)$ expansion necessary to restore positivity in the evolution equation. 
Since the pure NLO contribution to the width is negative, the crossing of the LO and NLO contribution at about 160~MeV signals that the added NNLO terms become as large as the pure NLO term.}
The feature of the AO widths becoming small at low temperatures may be phenomenologically relevant, as lattice QCD studies of the heavy-quark diffusion coefficient suggest a larger $\hat\kappa$ at lower temperatures. 
While this increase would tend to enhance the suppression, the reduction of the All-Orders width acts in the opposite direction and may partially compensate this effect. In contrast, applying the LO or NLO equations outside their regimes of validity could lead to an overestimation of the suppression.
Furthermore, we observe a gap between the full All-Orders width and the positive All-Orders widths that quantifies the impact of the negative sector on this observable. At high temperatures, this correction due to the negative sector amounts to $20\%-30\%$, while at low temperatures, where the width becomes very small, the correction exceeds $50\%$. 
The dashed line corresponds to the AO$+$ width with the Frobenius norm $\rho_\text{w}=\mathds{1}$, while we do not show the other norm choices because they practically coincide with the Frobenius-norm result.
Since the implemented operator optimization minimizes the negative contributions, the residual negative corrections to the in-medium width appear to be somewhat specific to the adopted correlator model. 
While parts of the negative contributions are expected to be unphysical, as they lead to negative eigenvalues in the density matrix, this is not necessarily true for the entire negative term. 
We therefore advocate treating the difference between full and positive width as a systematic uncertainty associated with the positivity-breaking sector, arising from the approximations used to derive the master equation, in particular the Born and Markov approximations. By means of illustration, we checked that optimizing for the $1\mathrm{S}$ width directly, by choosing $\rho_\text{w}=\ketbra{1\mathrm{S}}{1\mathrm{S}}$, significantly reduces the difference between the full and the positive $1\mathrm{S}$ width. 
However, such a weighting would lead to larger negative contributions in other parts of the Hilbert space, espcially the excited states.  

\subsection{Bjorken-evolution simulations}
\label{sec:bjorken}
To investigate the in-medium evolution of bottomonium using the All-Orders master equation, we solve Eq.~\eqref{eq:finalAO} under a Bjorken temperature evolution, 
\begin{equation}
    T(t) = T_0\left(\frac{t_0}{t}\right)^{1/3},
\end{equation}
with $t_0=0.6\,$fm, evolving from $T_0=400\,$MeV down to $155\,$MeV. 
Before turning on the medium evolution at $t_0$, we evolve the wave function from $t=0\,$fm to $t_0$ with the vacuum Hamiltonian. 
We note that this is the first investigation of bottomonium evolution using the QTraj code at temperatures below $190\,$MeV, as lower temperatures were previously excluded due to the $E/(\pi T)$ expansion. 

Similar to previous works~\cite{Brambilla:2022ynh,Brambilla:2023hkw,BR:2025lhx}, we consider two simulation modes: deterministic evolution with the effective Hamiltonian and full quantum evolution, including jump terms. 
The non-Hermitian effective Hamiltonian for a color state $u$ is given by
\begin{equation}
    H_{u,\text{eff}}= H_u-\frac{\mathrm{i}}{2}\sum_v \Gamma_{u\to v}.
    \label{eq:Heff}
\end{equation}
This is equivalent to solving the master equation~\eqref{eq:finalAO} by keeping
only the anticommutator term, while neglecting the jump term. 
For the full AO equation, the independence of the widths in Eqs.~\eqref{eq:singletwidth}-\eqref{eq:octetoctetwidth} from $\mu_J$ implies that the corresponding effective Hamiltonian is independent of this arbitrary scale as well. The $H_\text{eff}$ simulations can serve as a computationally inexpensive approximation for the solution of the master equation, as they do not require a stochastic unraveling, but rather can be obtained from a deterministic time evolution using~\eqref{eq:Heff}. 
This also applies in the presence of negative contributions in the case of the full master equation without increasing the computational cost. 
However, neglecting the jump term can overestimate the suppression, as it ignores possible recombination from higher-angular-momentum states or the octet sector that could contribute to the S-wave bottomonium overlap at later times. 
To account for this recombination, we consider full simulations of the AO$+$ Lindblad equation obtained by retaining only the positive sector of the optimally rotated pseudo-Lindblad equation.

In practice, we use QTraj~\cite{Omar:2021kra} to perform a standard unraveling of a Lindblad equation, see, e.g., Refs.~\cite{Omar:2021kra,Brambilla:2022ynh,Daley_2014}. In the case of $H_\text{eff}$ simulations, the evolution with the non-Hermitian Hamiltonian can be performed directly for the AO and AO$+$ cases.
To simulate the AO$+$ master equation including quantum jumps, we perform a standard stochastic unraveling, where we apply the optimal operator rotation according to Eqs.~\eqref{eq:phi} and~\eqref{eq:w} at each time step, using the Frobenius norm, i.e. $\rho_\text{w}=\mathds{1}$. 
To obtain the jump operators~\eqref{eq:jump1}-\eqref{eq:jump6}, we precompute ${A^{uv}}^\dagger_{l\pm1,l}$ for eleven different temperatures $T\in[0.15\,\text{GeV},0.65\,\text{GeV}]$ by solving the integral~\eqref{eq:A} numerically, using the correlator introduced in Sec.~\ref{sec:correlator}. 
The simulation results presented below correspond to the coupling fixed at the central scale $\mu=1.5\pi T_\text{ref}$. 
In this manner, we obtain ${A^{uv}}^\dagger_{l\pm1,l}$ for $l\leq 14$. 
If a trajectory jumps to $l>14$, we set its subsequent S-wave overlaps to zero, since such high-angular-momentum sectors have negligible probability to return to the S-wave sector. 
We checked that the S-wave survival probabilities are numerically converged with respect to this angular-momentum cutoff. 
In the QTraj code, we load in total $990$ sparse matrices representing the precomputed ${A^{uv}}^\dagger_{l\pm1,l}$ operators at the initialization. 
For temperatures that lie between the 11 precomputed temperatures, we linearly interpolate between two adjacent ${A^{uv}}^\dagger_{l\pm1,l}$ operators.
We then determine $\phi$ and $w$ from Eqs.~\eqref{eq:phi} and~\eqref{eq:w} for each pair of the jump operators~\eqref{eq:jump1}-\eqref{eq:jump6} and calculate the linear combination~\eqref{eq:operatorrotation}. 
If a trajectory is in a singlet state, the rotation only needs to be calculated for the two operator pairs $C^{\uparrow\pm}_{s\to o}$ and $C^{\downarrow\pm}_{s\to o}$. 
In contrast, if the trajectory is in an octet state, we perform the optimization for the four operator pairs $C^{\uparrow\pm}_{o\to s}$, $C^{\downarrow\pm}_{o\to s}$, $C^{\uparrow\pm}_{o\to o}$, and $C^{\downarrow\pm}_{o\to o}$.
We perform the simulation at a box size of $L=40\,\rm{GeV}^{-1}$, discretized on an $N=1500$ lattice. 
Following Ref.~\cite{Brambilla:2022ynh}, for the bottom-quark mass we choose $M=4.73\,$GeV and set the coupling of the potential to $\alpha_s=0.468$.
As initial state, we choose a color-singlet state with $l=0$ and a compact Gaussian wave function in radial space. 
The rest of the computational setup is kept identical to previous studies, for which we refer the reader to Refs.~\cite{Omar:2021kra,Brambilla:2022ynh}. 

We compare our simulation results of the All-Orders master equation against simulations of the LO and NLO master equations using QTraj. 
Here, we keep all simulation parameters equal to the All-Orders simulations and set $\hat\kappa=2.5$ and $\hat\gamma=-1.55$ to ensure an equal comparison with our spectral function model. 
We note that the NLO master equation derived in Ref.~\cite{Brambilla:2022ynh} does not contain $\tilde\gamma$ due to the instantaneous approximation made in the derivation. 
We therefore perform the NLO simulations setting $\tilde\gamma=0$. 
Since the AO evolution retains a nonzero $\tilde{\gamma}$
generated by our spectral model, the reported differences between the AO and NLO equations are not exclusively due to higher orders in $E/(\pi T)$, but also contain the effect of the instantaneous approximation used in the derivation of the NLO equation.

\begin{figure}
    \centering
    \includegraphics[width=0.49\linewidth]{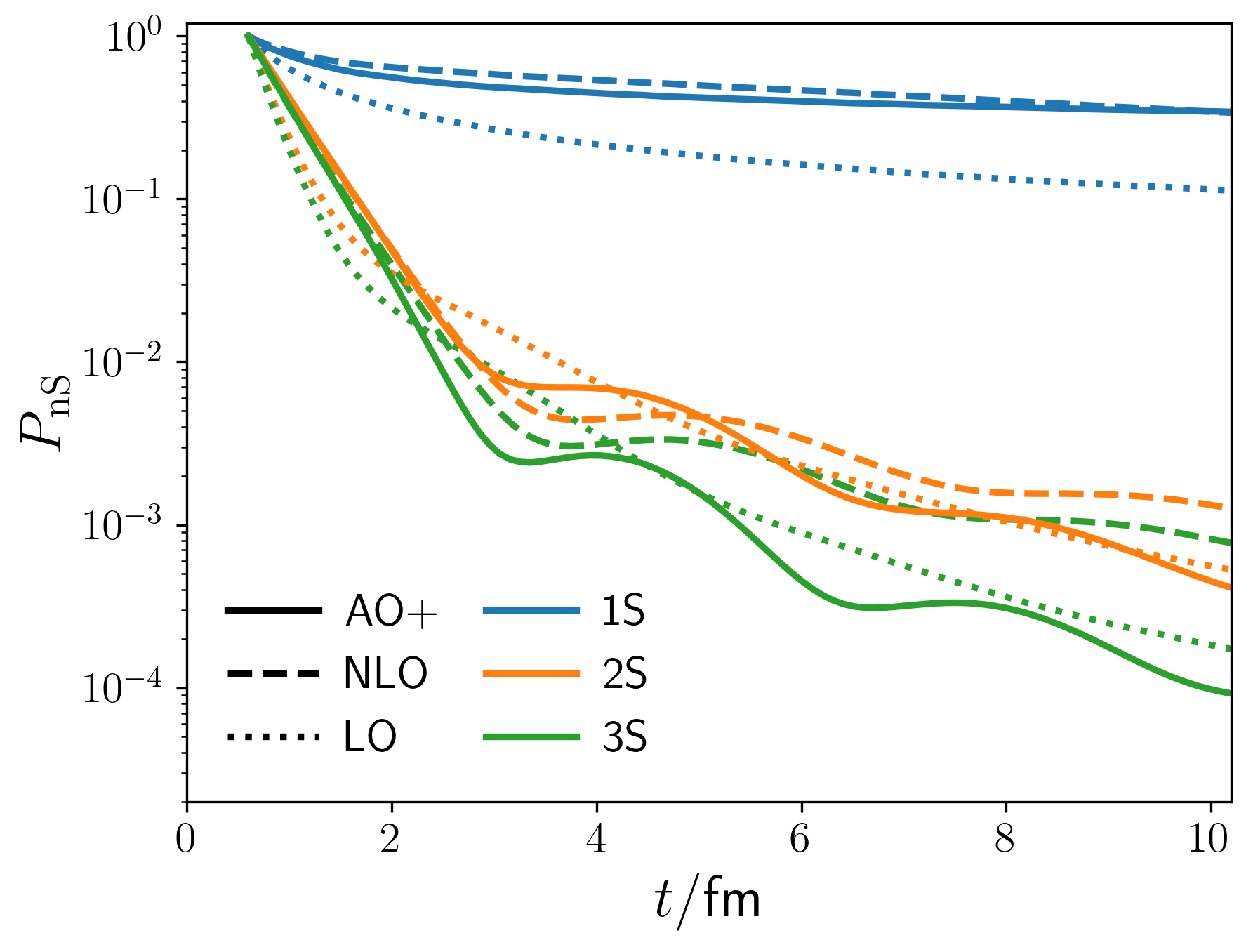}
    \includegraphics[width=0.49\linewidth]{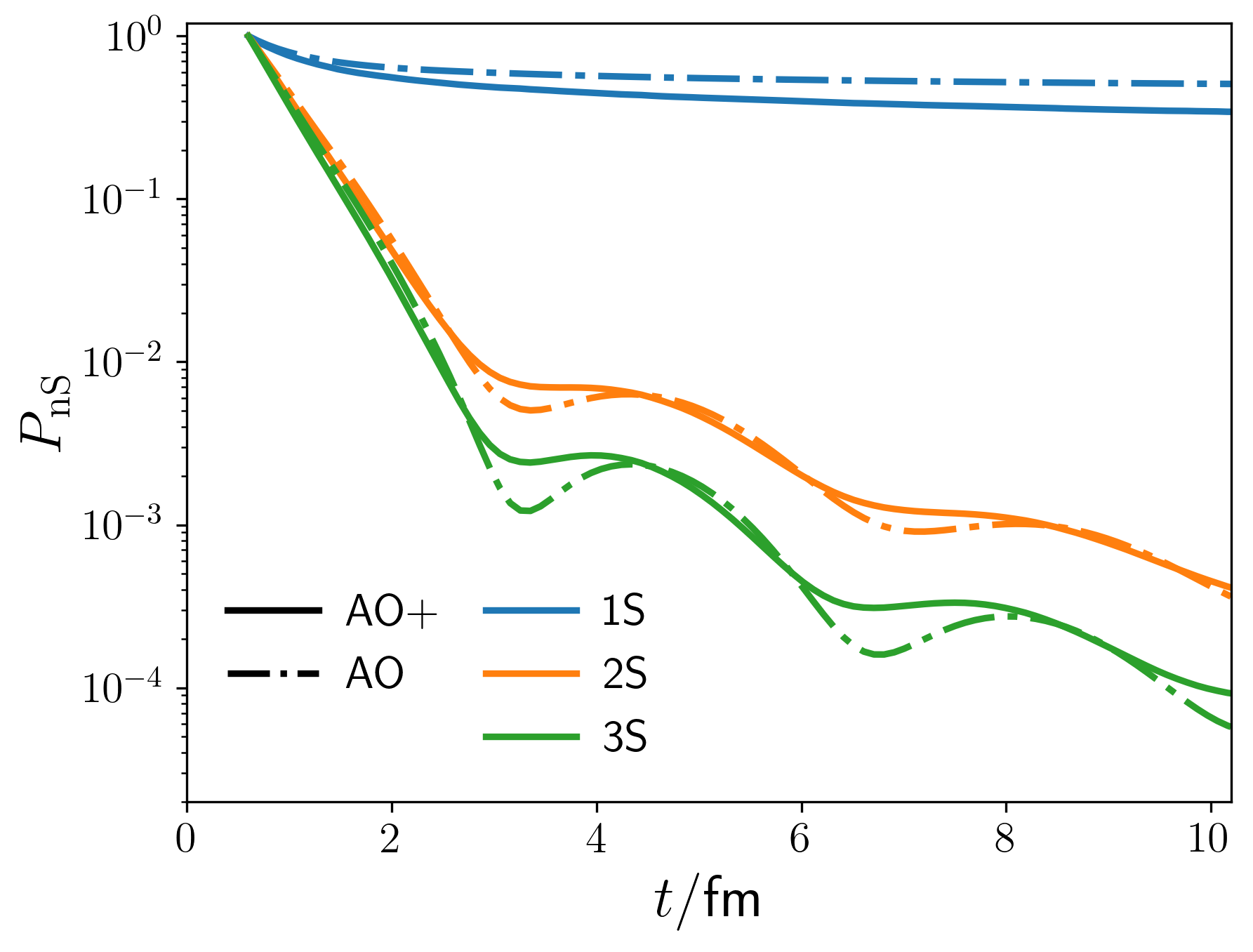}
    \caption{Simulation results of the survival probabilities $P_\text{nS}$ for the $1\mathrm{S},2\mathrm{S}$ and $3\mathrm{S}$ states with the spectral function model from Sec.~\ref{sec:correlator} using the evolution with the effective Hamiltonian $H_\text{eff}$, excluding quantum jumps.
    \textit{Left:} We compare the positive All-Orders master equation (solid) with the NLO (dashed) and LO (dotted) master equations. 
    \textit{Right:} We compare the positive All-Orders master equation (solid) with the full All-Orders master equation including negative terms (dash-dotted).}
    \label{fig:bjorken1}
\end{figure}

In the left panel of Fig.~\ref{fig:bjorken1}, we show the time evolution of the $\Upsilon(1\mathrm{S})$, $\Upsilon(2\mathrm{S})$, $\Upsilon(3\mathrm{S})$ survival probabilities, 
\begin{align}
P_\mathrm{nS}(t) = \frac{\text{Tr}\left[\rho(t)\ketbra{\mathrm{nS}}{\mathrm{nS}}\right]}{\text{Tr}\left[\rho(0)\ketbra{\mathrm{nS}}{\mathrm{nS}}\right]},
\end{align}
under $H_\text{eff}$ evolution of the AO$+$ master equation, where $\ket{\mathrm{nS}}$ is the $\mathrm{n}$-th eigenstate of the vacuum singlet Hamiltonian with zero orbital angular momentum. 
We compare the AO$+$ simulation results with the $H_\text{eff}$ simulations of the LO and NLO master equations. 
We observe that the AO$+$ master equation leads to slightly stronger suppression of the $1\mathrm{S}$ at the beginning of the evolution compared to the NLO curve, consistent with the width results in Fig.~\ref{fig:width}. 
At later times, the $1\mathrm{S}$ survival probability plateaus for the AO$+$ case, while for the NLO equation it continues to decrease, in accordance with the findings in~\cite{BR:2025lhx}. 
For the excited states, the differences between AO+, NLO, and LO are smaller during the early, high-temperature part of the evolution. 
This is consistent with the excited states having a smaller binding energy, so that the $E/(\pi T)$ expansion behaves better during the early, high-temperature part of the evolution.
In the right panel of Fig.~\ref{fig:bjorken1}, we compare the $H_\text{eff}$ evolution of the positive All-Orders master equation with that of the full All-Orders equation, including the negative terms. 
We notice that the $1\mathrm{S}$ survival probability receives small corrections from the negative sector, with the full All-Orders evolution plateauing faster than the positive one, in accordance with the width comparison in Fig.~\ref{fig:width}.
Furthermore, the $2\mathrm{S}$ and $3\mathrm{S}$ survival probabilities show minor corrections.

\begin{figure}
    \centering
    \includegraphics[width=0.49\linewidth]{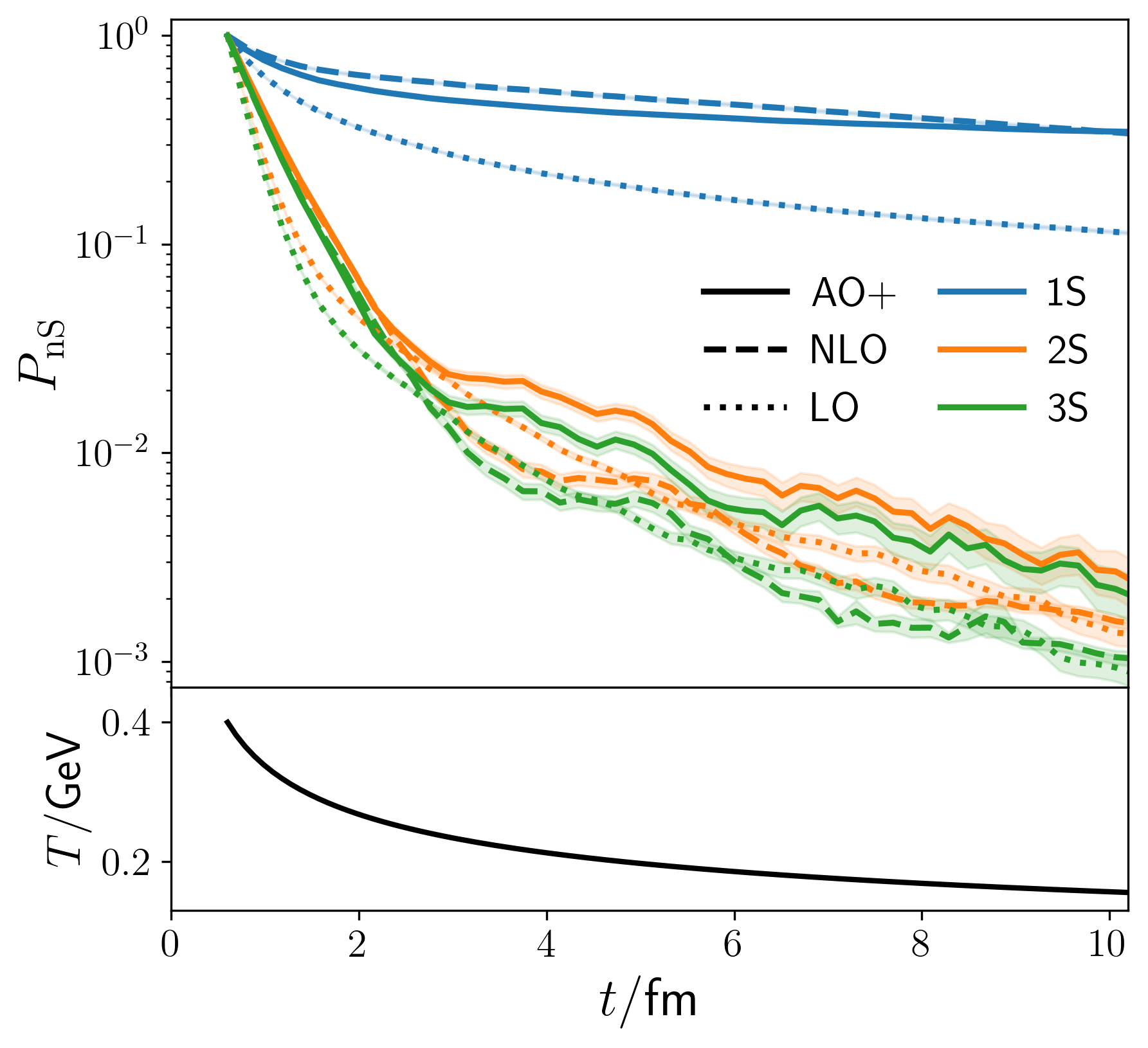}
    \includegraphics[width=0.49\linewidth]{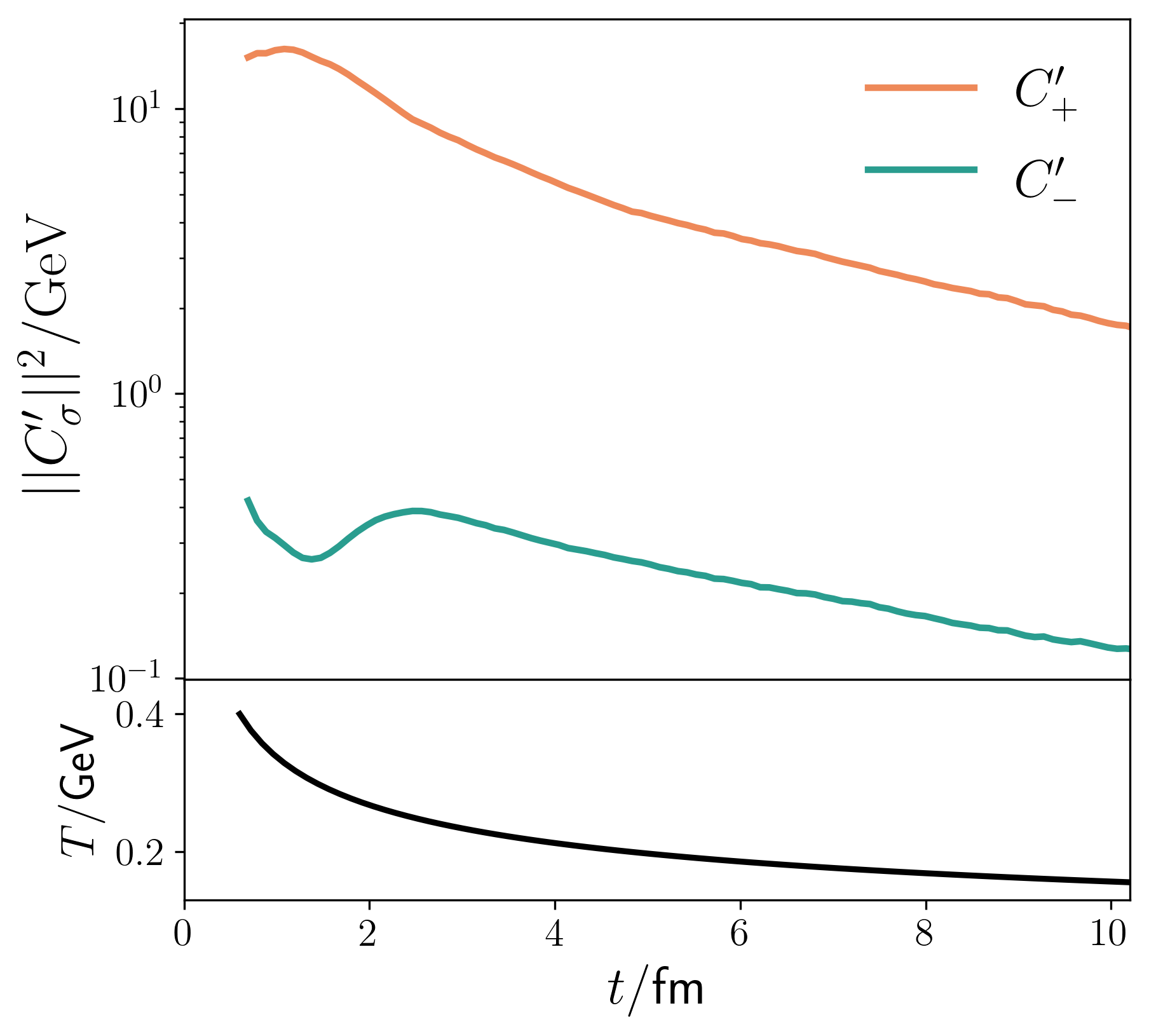}
    \caption{\textit{Left:} Simulation results of the survival probabilities $P_\mathrm{nS}$ for the $1\mathrm{S},2\mathrm{S}$ and $3\mathrm{S}$ states including quantum jumps. We compare the positive All-Orders master equation (solid) with the NLO (dashed) and LO (dotted) master equations. 
    The shaded bands represent the statistical uncertainty from averaging over stochastic trajectories. 
    \textit{Right:} The time evolution of the average optimized positive- and negative-sector squared Frobenius norms of the AO$+$ simulation.}
    \label{fig:bjorken2}
\end{figure}

Finally, in the left panel of Fig.~\ref{fig:bjorken2}, we show the evolution of the survival probabilities from quantum-jump simulations. 
We do not include the full All-Orders master equation because simulations at these lattice sizes, including quantum jumps, are computationally too expensive when the negative sector is included. 
We comment on the simulation of the full All-Orders master equation, including quantum jumps, in Appendix~\ref{app:plqt}. 
We observe that the $1\mathrm{S}$ survival probabilities do not receive sizable corrections from the quantum jumps, quantitatively agreeing with the $H_\text{eff}$-only case.
The $2\mathrm{S}$ and $3\mathrm{S}$ survival probabilities, on the other hand, receive non-negligible corrections from quantum jumps, leading to less suppression due to regeneration. 
Furthermore, we observe that for the excited states, the positive All-Orders master equation agrees well with the NLO master equation at high temperatures, while it leads to less suppression at low temperatures. 

Since these results are obtained in the AO$+$ positive-sector approximation, we check how large the optimized positive sector and the residual negative sector are throughout the evolution. 
In the right panel of Fig.~\ref{fig:bjorken2}, we show the average squared Frobenius norm of the optimized positive- and negative-sector jump operators as a function of time. 
For each trajectory and time step, we compute the squared Frobenius norms of all positive- and negative-sector jump operators available from the trajectory's current color and angular-momentum sector, average these norms separately over the positive and negative sectors, and then average the result over all trajectories. 
Since these norms are evaluated after the optimal operator rotation, they are likewise independent of the normalization scale $\mu_J$. 
We see a clear separation between the positive and negative operator norms of about an order of magnitude, showing that the operator optimization establishes a hierarchy between the retained positive sector and the discarded negative sector. 
The hierarchy persists at later times, which are associated with lower temperatures, supporting the effectiveness of the operator optimization and enabling us to extend the evolution into the low-temperature regime.

\section{Conclusion and outlook}
\label{sec:conclusions}
The central result of this work is a framework that couples the real-time evolution of bottomonium in the quark-gluon plasma directly to suitable convolutions of Coulombic Green's functions with chromoelectric correlators, removing the restriction that the medium dependence be encoded only in few transport coefficients. By avoiding to expand in the binding energy over temperature, while enabling a quantum-trajectory-based simulation, we extend the open-quantum-system description of bottomonium suppression across the entire temperature range relevant to heavy-ion collisions, including the low-temperature regime below $190\,$MeV that was previously inaccessible.

We project the original master equation onto spherical harmonics, obtaining the All-Orders master equation, consisting of 12 jump operators. 
Six of them enter the master equation with a negative sign, breaking positivity and thus leading to a master equation that is not of the Lindblad type.
The resulting master equation can, in principle, be simulated with a stochastic unraveling scheme. 
However, while simulations with $H_\text{eff}$ that neglect quantum jumps can be performed efficiently, a full simulation requires substantially more computational resources than in the Lindblad case, making large-scale simulations impractical. 
To enable simulations that include quantum jumps, we take a different route. 
Assuming that the approximations made in the derivation of the master equation are justified, we expect the negative contributions to be small. 
Under this assumption, we utilize the concept of optimal operator rotations to select a basis for the jump operators that minimizes the norm of the negative jump operators. 
We then retain only the positive sector of the optimized representation, obtaining a Lindblad equation, which can be efficiently unraveled using a standard Monte Carlo wave function approach.

To perform the simulations in practice, we build a model for the spectral function of the chromo-electric correlator that smoothly interpolates between the known IR and UV limits. 
Using the obtained correlator, we compute the in-medium widths of the 1S bottomonium with and without the negative terms. 
At low temperatures, the resulting width decreases rapidly, while the width at the next-to-leading-order in $E/(\pi T)$ remains notably larger. 
This behavior may be phenomenologically important, since the reduction of the All-Orders width acts in the opposite direction to the expected increase of $\hat\kappa$ toward lower temperatures and may therefore partially compensate the associated enhancement of the suppression. Different choices of the norm in the operator optimization do not have a significant impact. 

Simulating the evolution of the bottomonium in a Bjorken medium, we find that the $\Upsilon(1\mathrm{S})$ survival probability is described well by $H_\text{eff}$ simulations, while the excited states require full simulations including quantum jumps. 
Comparing the All-Orders and NLO simulations, we find that the $\Upsilon(1\mathrm{S})$ survival probability shows slightly more suppression in the AO$+$ case, as expected from the in-medium width calculation. 
Since quantum jumps have a negligible effect on the $\Upsilon(1\mathrm{S})$ survival probability, future phenomenological studies can estimate the systematic uncertainty associated with discarding the negative sector by comparing the computationally inexpensive $H_\mathrm{eff}$ evolutions corresponding to the AO$+$ and full AO master equations, without requiring a full stochastic unraveling of the latter. 
In the $H_\text{eff}$ simulations, the $\Upsilon(2\mathrm{S})$ and $\Upsilon(3\mathrm{S})$ survival probabilities show only very small corrections between the AO$+$ and AO master equations, suggesting that the positive sector approximation is well behaved for the excited states. 
Comparing the full simulations of the AO$+$ master equation, including quantum jumps, with the NLO master equation, the excited states agree well at high temperatures, while at low temperatures, the All-Orders master equation leads to less suppression. The presented results enable the extension of open-quantum-system-based simulations of bottomonium suppression to temperatures below $190\,$MeV, thereby allowing the treatment of the entire temperature range relevant to heavy-ion collisions. 

A natural next step will be to study the phenomenology of bottomonium suppression, including hydrodynamics and feeddown, using this framework. 
The key input to this program is the real-time chromoelectric correlator, which fully characterizes the color medium beyond a finite set of transport coefficients. 
Standard lattice-QCD calculations determine correlators in Euclidean time, and extracting the corresponding real-time correlator requires solving an ill-posed analytic-continuation problem, commonly referred to as the lattice inverse problem. 
One possible avenue for constraining the correlator is to combine lattice data with theoretical information, for example, within a Bayesian analysis. 
Complementary real-time approaches, including classical-statistical lattice simulations~\cite{Boguslavski:2018beu}, complex-Langevin simulations~\cite{Boguslavski:2022dee}, Hamiltonian and tensor-network techniques~\cite{Banuls:2015sta}, and quantum simulations of lattice gauge theories~\cite{Martinez:2016yna,Bauer:2022hpo}, may provide additional information.
Supplying real-time correlators as input to the All-Orders master equation derived here would connect non-perturbative real-time QCD directly to quarkonium-suppression phenomenology.

Finally, this work paves the way for further refinements of the open quantum system formulation.
A first step beyond the Markov approximation would be to compute the operators in Eq.~\eqref{eq:Aop} over a finite upper integration time, yielding time-dependent operators.
In the future, a full phenomenological study should also consider higher-order perturbative corrections to the Coulomb potential~\cite{Brambilla:2024tqg} in the All-Orders master equation, which are expected to affect excited states.

\section*{Acknowledgments}
T.M. thanks Alexander Rothkopf for providing the in-medium width data. 
N.B., T.M., and A.V. thank Panayiotis Panayiotou for useful discussions. 
N.B., T.M., and A.V. acknowledge support by the DFG cluster of excellence ORIGINS funded by the Deutsche Forschungsgemeinschaft (DFG) under Germany's Excellence Strategy - EXC-2094-390783311. The work of N.B. is supported by the DFG Grant No. BR 4058/5-1 ``Open quantum systems and effective field theories for hard probes of hot and/or dense medium”. 
N.B. acknowledges the advanced grant ERC-2023-ADG-Project EFT-XYZ. 
The authors gratefully acknowledge the Gauss Centre for Supercomputing e.V. (\href{https://www.gauss-centre.eu/}{www.gauss-centre.eu}) for funding this project by providing computing time on the GCS Supercomputer SuperMUC-NG at the Leibniz Supercomputing Centre (\href{https://www.lrz.de/}{www.lrz.de}). 

\appendix
\begin{appendix}

\section{Spherical basis projection} 
\label{app:angularprojection}
Scalar operators of the form $S_1  r_i S_2 r_i S_3$, in which $S_1, S_2, S_3$ are scalar operators and $r_i$ is the position operator, appear when projecting the evolution equations \eqref{eq:general_singlet} and \eqref{eq:general_octet} onto the angular momentum eigenstates, which are spherical harmonics in position space. 
In this appendix, we show in detail the procedure of projecting $S_1  r_i S_2 r_i S_3$ onto spherical harmonics. 

First, the density matrix $\rho$, as well as any scalar operator $S$, are rotationally invariant and therefore block diagonal in the angular momentum quantum number $l$ and degenerate in the magnetic quantum number $m$. 
We denote $\rho^l = \sum_{m}\bra{lm} \rho \ket{lm} = (2l+1) \bra{lm} \rho \ket{lm} $, following the main text and accounting for degeneracy in $m$, and $S^l = \bra{lm} S \ket{lm}$ for any other scalar operator $S$. 
To initiate the projection, we insert a complete set of angular momentum eigenstates
\begin{align}
\bra{lm}S_1  r_i S_2 r_i S_3 \ket{lm} =  S_1^l \left(\sum_{l'm'}\bra{lm} r_i \ket{l'm'} S^{l'}_2 \bra{l'm'}r_i \ket{lm} \right)  S_3^l .  \label{eq: scalar op proj}
\end{align} 
Thus, Eq.~\eqref{eq: scalar op proj} reduces to evaluating the matrix elements of the position operator $\bra{l'm'}r_i \ket{lm}$. 
In spherical coordinates $(r, \theta,\phi)$, the angular momentum eigenstates are spherical harmonics
\begin{align}
\braket{\theta,\phi | lm} = Y_{lm} (\theta,\phi) = (-1)^m \sqrt{\frac{2l+1}{4\pi} \frac{(l-m)!}{(l+m)!}} P^m_l(\cos\theta) e^{im\phi} = N_{lm} P^m_l(\cos\theta) e^{im\phi}, 
\end{align}
and the Cartesian coordinates $r_i$ are related to spherical coordinates by 
\begin{align}
 x + \mathrm{i} y = r \sin\theta \,\mathrm{e}^{\mathrm{i}\phi},\qquad 
 (x-\mathrm{i}y) = r \sin\theta \,\mathrm{e}^{-\mathrm{i}\phi},\qquad  
 z = r \cos\theta. 
\end{align}
By applying recurrence relations of the associated Legendre polynomials \begin{align} 
\begin{split}
x P^m_l(x)  & = \frac{1}{2l+1} \Big[
(l+m) P^m_{l-1}(x) + (l-m+1) P^m_{l+1}(x) \Big] ,  \\
\sqrt{1-x^2} \, P^m_l(x) & =\frac{1}{2l+1} \Big[
P^{m+1}_{l+1}(x) - P^{m+1}_{l-1}(x)
\Big],
\end{split}
\end{align} 
as well as the orthogonality condition
\begin{align}
 \int_{-1}^{1} P^m_l(x) P^m_{l'} (x) \dd x  = \frac{2}{(2l+1)} \frac{(l+m)!}{(l-m)!} \delta_{l,l'}
\end{align} 
to $P^m_l(\cos\theta)$, we obtain \begin{align}
\bra{l'm'} (x+\mathrm{i}y) \ket{lm} = r C_+, \qquad
\bra{l'm'} (x-\mathrm{i}y) \ket{lm}  = r C_-, \qquad 
\bra{l'm'} z \ket{lm}   = r C_z,
\end{align}
where we have defined 
\begin{align}
C_+ &=      \delta_{m',m+1} \left( -
\delta_{l',l+1} \sqrt{ \frac{l+m+1}{2l+1} \frac{l+m+2}{2l+3}   }  + \delta_{l',l-1} \sqrt{
\frac{l-m-1}{2l-1} \frac{l-m}{2l+1} 
} 
\right),  \\
C_- &=  \delta_{m',m-1} \left( \delta_{l',l+1} \sqrt{\frac{l-m+1}{2l+1} \frac{l-m+2}{2l+3}} - \delta_{l',l-1} \sqrt{\frac{l+m-1}{2l-1} \frac{l+m}{2l+1}}
\right),  \\
C_z & = \delta_{m',m} \left(\delta_{l',l+1} \sqrt{ \frac{l-m+1}{2l+1}
\frac{l+m+1}{2l+3}
}+ 
\delta_{l',l-1} \sqrt{ \frac{l-m}{2l-1}
\frac{l+m}{2l+1}
}    
\right). 
\end{align}
Then, the dot product in Cartesian coordinates can be expanded \begin{align}
\begin{split}
 & \bra{lm} r_i \ket{l'm'} S^{l'}_2 \bra{l'm'}r_i \ket{lm}  \\
 & = \frac{1}{2}    \Big[
\bra{lm} x-\mathrm{i}y \ket{l'm'} S^{l'}_2  \bra{l'm'}x+\mathrm{i}y\ket{lm}   +  \bra{lm} x+\mathrm{i}y \ket{l'm'} S^{l'}_2   \bra{l'm'}x-\mathrm{i}y\ket{lm}  
\Big]  \\
& \qquad + \bra{lm} z \ket{l'm'} S^{l'}_2    \bra{l'm'}z\ket{lm}   \\
& = \frac{1}{2}    \Big[
\bra{l'm'} x+\mathrm{i}y \ket{lm} S^{l'}_2    \bra{l'm'}x+\mathrm{i}y\ket{lm} + \bra{l'm'} x-\mathrm{i}y \ket{lm} S^{l'}_2   \bra{l'm'}x-iy\ket{lm}  
\Big] \\
& \qquad + \bra{l'm'} z \ket{lm} S^{l'}_2   \bra{l'm'}z\ket{lm}   \\
& = \left(\frac{1}{2} (C_+^2+C_-^2)+ C_z^2 \right)
r S^{l'}_2   r   . 
\end{split}
\label{eq: selection rule}
\end{align} 
Summing over $m'$ we have \begin{align}
\sum_{m'} \left(\frac{1}{2} (C_+^2+C_-^2) +C_z^2\right) = \delta_{l', l+1} 
\frac{l+1}{2l+1} 
  + \delta_{l', l-1} \frac{l}{2l+1}. \label{eq: selection coefficients}
\end{align}
Finally, by substituting Eqs. (\ref{eq: selection rule}) and (\ref{eq: selection coefficients}) into Eq. (\ref{eq: scalar op proj}), we obtain the scalar operator projection formula  
\begin{align}
\bra{lm} S_1 r_i  S_2 r_i S_3 \ket{lm}   = S_1^l r \left(\frac{l+1}{2l+1}
 S_2^{l+1}       + \frac{l}{2l+1}  S_2^{l-1}\right)  r S_3^l . \label{eq: op proj formula}
\end{align}

\section{Agreement with NLO expansion}
\label{app:NLO expansion}
In this appendix, we peform the $E/(\pi T)$ expansion to NLO in the master equation and width operators in the reduced spherical representation \eqref{eq:off_diagonal_Lindblad}--\eqref{eq: all order width}. 
We show that, after imposing the additional instantaneous approximation $\Tilde{\gamma} =0$ used in Ref.~\cite{Brambilla:2022ynh}, we get a master equation and width operators equal to those derived in Ref.~\cite{Brambilla:2022ynh}, up to differences of $\mathcal{O}(E^2/(\pi T)^2)$. 
We recall that the LO and NLO width operators for singlet to octet transitions are given by
\begin{align}
    \Gamma_{s\to o}^{\text{LO}} &= \kappa r^2,\label{eq: Gamma so LO}\\
    \Gamma_{s\to o}^{\text{NLO}} &= \kappa \left[r^2\left(1+\frac{\Delta V_{so}}{4T}\right)^2 - \frac{3}{2MT} -\frac{\Delta V_{so}}{4MT^2} + \frac{\mathcal{D}^2}{4M^2T^2}\right], \label{eq: Gamma so NLO}
\end{align}
where 
\begin{equation}
    \mathcal{D}^2 = -\frac{\partial^2}{\partial r^2} + \frac{l(l+1)}{r^2},
\end{equation}
in the reduced spherical representation. 
The width operators~\eqref{eq: Gamma so LO} and~\eqref{eq: Gamma so NLO} are responsible for the 1S in-medium width at the specified order. The NLO width operator~\eqref{eq: Gamma so NLO} includes terms of $\mathcal{O}(E^2/(\pi T)^2)$ that are necessary to ensure its positivity. 
Our goal is to recover Eq.~\eqref{eq: Gamma so NLO}, and Eq.~\eqref{eq: Hamiltonian NLO}, up to $\mathcal{O}(E^2/(\pi T)^2)$ differences, by expanding Eqs.~\eqref{eq: Hamiltonian} and~\eqref{eq: all order width}. 

First, we expand the ${A^{uv}}_{l'l}$ operators and their conjugates under the reduced spherical representation while making use of Eq.~\eqref{eq:cond1} and Eq.~\eqref{eq:cond2}, 
\begin{align}
{A^{uv}}_{l'l} 
& =  \frac{\kappa - \mathrm{i}\gamma}{2} r + \left(\frac{\kappa}{4T} - \mathrm{i} \Tilde{\gamma}\right) (h_u^{l'} r - r h_v^l )+ \mathcal{O}(E^2/(\pi T)^2) \nonumber \\
& = \frac{\kappa- \mathrm{i}\gamma}{2} r + \left(\frac{\kappa}{4T} - \mathrm{i} \Tilde{\gamma} \right) \left(
- \frac{2}{M} \frac{\partial}{\partial r} + \frac{(l'+l+1)(l'-l)}{Mr} + r \Delta V_{uv}
\right) + \mathcal{O}(E^2/(\pi T)^2), \label{eq:NLOA} \\
{{A^{uv}}^\dag}_{l'l}  & = \frac{\kappa + \mathrm{i} \gamma}{2} r + \left(\frac{\kappa}{4T} + \mathrm{i} \Tilde{\gamma}\right) \left( r h_u^l - h_{v}^{l'} r\right) + \mathcal{O}(E^2/(\pi T)^2) \nonumber \\
& = \frac{\kappa+ \mathrm{i}\gamma}{2} r + \left(\frac{\kappa}{4T} + \mathrm{i} \Tilde{\gamma} \right) \left(
 \frac{2}{M} \frac{\partial}{\partial r} -\frac{(l'+l+1)(l'-l)}{Mr} + r \Delta V_{uv}
\right) + \mathcal{O}(E^2/(\pi T)^2) . \label{eq:NLOAdag}
\end{align}
Then, substituting Eqs.~\eqref{eq:NLOA} and~\eqref{eq:NLOAdag} into Eqs.~\eqref{eq: Hamiltonian} and~\eqref{eq: all order width}, we obtain the Hamiltonian at NLO
\begin{equation}
\begin{aligned}
H^l_u &= - \frac{1}{M} \frac{\partial^2}{\partial r^2} + \frac{l(l+1)}{Mr^2} + V_u(r)   \\
&\quad +  \sum_v c_{u \to v} \left[ \frac{\gamma}{2} r^2 - \mathrm{i} \frac{\kappa}{4MT} \left\{r , \partial_r \right\} 
+ \Tilde{\gamma} \left(-\frac{3}{M}+r^2 \Delta V_{uv}\right)
\right] + \mathcal{O}(E^2/(\pi T)^2),  
\end{aligned}
\label{eq:Hamiltonian expanded}
\end{equation}
as well as the decay width operators at NLO 
\begin{align}
\Gamma^{\uparrow}_{u\to v}  
& = c_{u\to v} N^\uparrow  \left[ \kappa   \left( \left( 1 + \frac{\Delta V_{uv}}{2T}\right) r^2 - \frac{2l+3}{2MT}  \right) + \Tilde{\gamma} \frac{2i}{M} \{r, \partial_r\}\right]+ \mathcal{O}(E^2/(\pi T)^2), \label{eq:width expanded up} \\
\Gamma^{\downarrow}_{u\to v}  
& =  c_{u\to v}N^\downarrow \left[ \kappa   \left(  \left( 1 + \frac{\Delta V_{uv}}{2T}\right) r^2 + \frac{2l-1}{2MT} \right) + \Tilde{\gamma} \frac{2i}{M} \{r, \partial_r\}\right]+ \mathcal{O}(E^2/(\pi T)^2). \label{eq:width expanded down}
\end{align}
Following the instantaneous approximation in Ref.~\cite{Brambilla:2022ynh}, we set $\Tilde{\gamma} =0$. 
After expanding to NLO in $E/(\pi T)$ and setting $\Tilde{\gamma}=0$, Eqs.~~\eqref{eq:Hamiltonian expanded}--\eqref{eq:width expanded down}, indeed, only differ from the Hamiltonian and width operators given in Ref. \cite{Brambilla:2022ynh} by  $\mathcal{O}(E^2/(\pi T)^2)$ terms. 
To illustrate the point, we present the summed singlet width $\Gamma_{s\to o}$: \begin{align}
\Gamma_{s\to o} = \kappa \left[
\left(1 + \frac{\Delta V_{so}}{2T}\right) r^2  
- \frac{3}{2MT}
\right] + \mathcal{O}(E^2/(\pi T)^2).  
\label{eq:Gammaexpanded}
\end{align}
It differs from Eq.~\eqref{eq: Gamma so NLO} by $\mathcal{O}(E^2/(\pi T)^2)$ terms.  
The $\mathcal{O}(E^2/(\pi T)^2)$ difference between Eq.~\eqref{eq:Gammaexpanded} and Eq.~\eqref{eq: Gamma so NLO} arises from the operator rotation performed after the NLO expansion. 
While Eq.~\eqref{eq:Gammaexpanded} comes directly from performing the NLO expansion, Eq.~\eqref{eq: Gamma so NLO} contains a positivity-restoring completion that ensures that the evolution of the trajectories remains physical.

\section{Spectral function choice}
\label{app:spec}
While the model for the spectral function presented in Sec.~\ref{sec:correlator} respects the correct IR and UV behavior, it is not a unique choice. In this Appendix, we illustrate how the presented results depend on the choice of the switch function $s(\omega)$. 
To do so, we inspect the dependence of the in-medium width and the chromoelectric memory timescale on the parameters $\Delta = \chi = (\pi/2)\,T$, by varying them by $\pm 25\%$. 
We find that the width is insensitive to this variation, with sub-percent deviations.
\begin{figure}[h]
    \centering
    \includegraphics[width=0.49\linewidth]{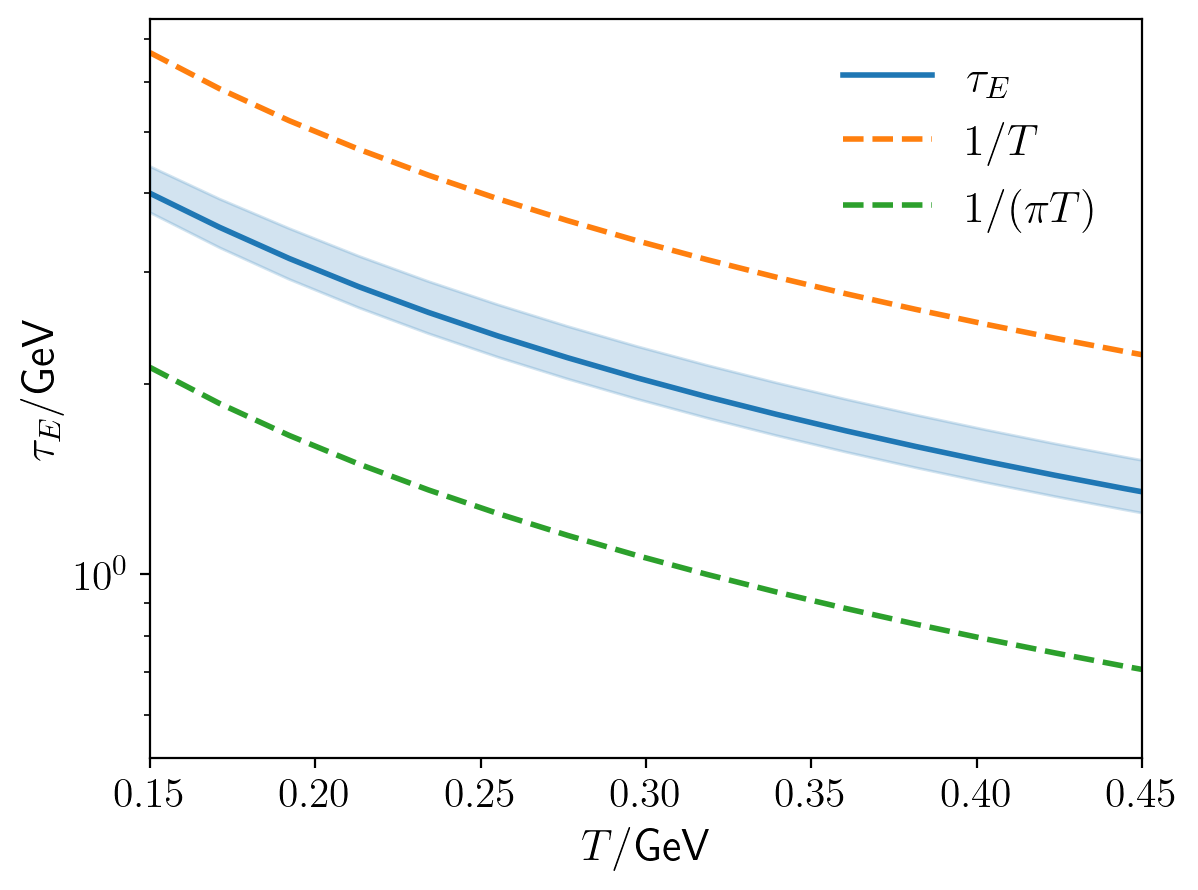}
    \includegraphics[width=0.49\linewidth]{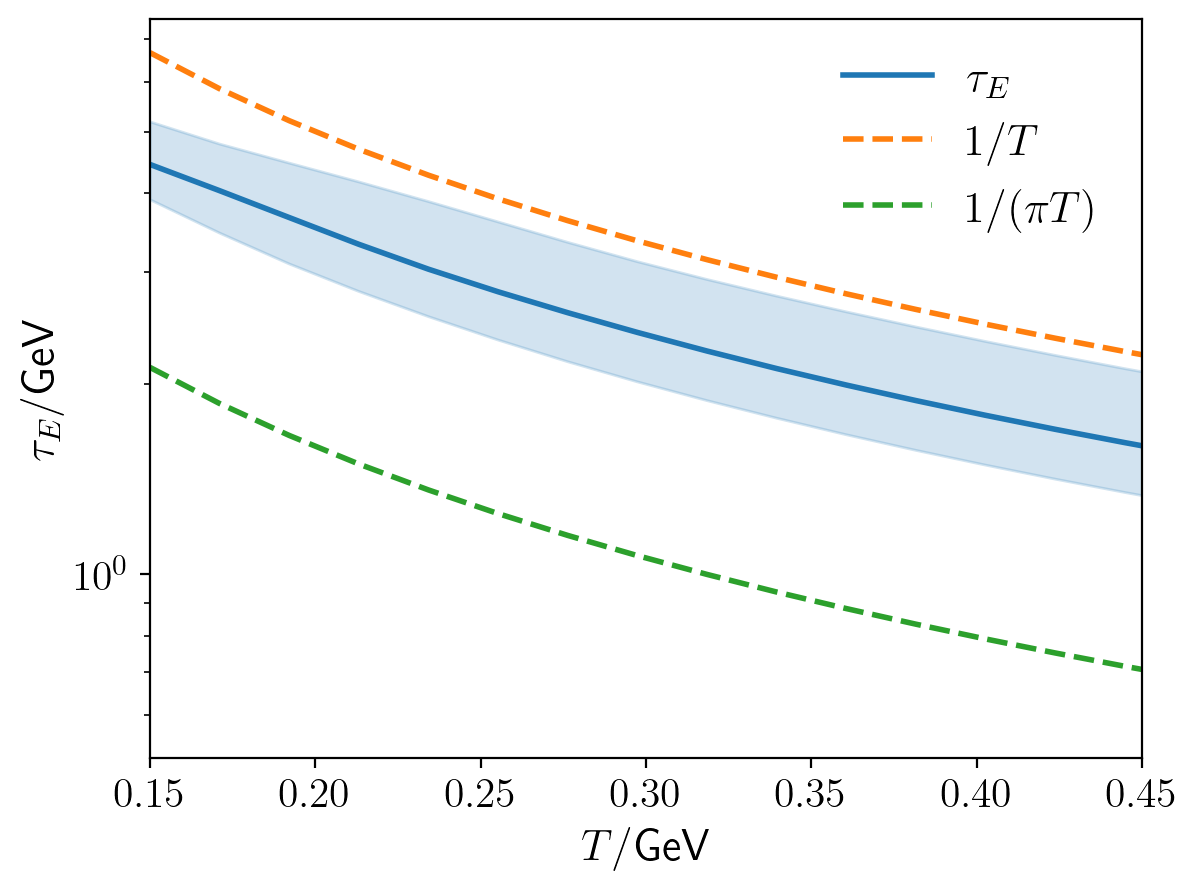}\caption{
    \textit{Left:} The temperature dependence of the chromoelectric memory timescale $\tau_E$ for the switch function defined in Eq.~\eqref{eq:tanhansatz}. 
    The blue shaded band corresponds to a $\pm 25\%$ variation of the spectral function model parameters $\Delta$ and $\chi$. 
    The coupling $g^2$ is fixed at the central scale $\mu=1.5\pi T_\text{ref}$.
    \textit{Right:} The chromoelectric timescale $\tau_E$ for the Lorentzian switch function ansatz. 
    The coupling $g^2$ is again fixed at the central scale $\mu=1.5\pi T_\text{ref}$, while the band corresponds to a $\pm 25\%$ variation of the parameter $\Lambda$.}
    \label{fig:tanhband}
\end{figure}
This is expected, since in the $E/(\pi T)$ expansion, the width depends only on $\hat\kappa$; at fixed $\hat\kappa$, the dependence on the shape of the spectral function enters only through higher-order corrections in the $E/(\pi T)$ expansion. By contrast, the chromoelectric memory timescale shown in Fig.~\ref{fig:tanhband} exhibits a stronger dependence than the in-medium width, but remains well within the range between $1/T$ and $1/(\pi T)$. We find the strongest dependence on  $\Delta$ and $\chi$ for the transport coefficient $\hat\gamma$. Calculating $\hat\gamma$ for a $\pm 25\%$ variation of the parameters, we find $\hat\gamma$ to vary between $-1.3$ and $-1.8$.

In addition to the aforementioned model, we investigate a second model for the switch function $s(\omega)$ given by a simple Lorentzian 
\begin{equation}
    s(\omega) = \frac{\omega^4}{\Lambda^4 + \omega^4}.
\end{equation}
We choose $\Lambda=(\pi/2)\,T$ again and vary the value by $\pm 25\%$. We find that the in-medium width agrees with the results calculated from the switch function defined in Eq.~\eqref{eq:tanhansatz} within sub-percent accuracy, with negligible dependence on the parameter $\Lambda$. 
The resulting memory timescale for this spectral function is shown in Fig.~\ref{fig:tanhband}. To confirm the robustness of the results across a broader class of spectral functions, a systematic study of the spectral-function dependence, ideally relying on 
further input from lattice QCD, is needed. 
We defer such studies to future work.

\section{Pseudo-Lindblad Quantum Trajectories for the full master equation}
\label{app:plqt}
In this Appendix, we provide details on the unraveling of the full master equation, including the negative terms. 
While there are different unraveling schemes, we focus on the Pseudo-Lindblad Quantum Trajectory (PLQT) algorithm proposed in Ref.~\cite{Becker_2023}.
The PLQT algorithm enables the solution of a master equation with jump operators $L_i$, whose terms enter with prefactors $\gamma_i$, specifically allowing for $\gamma_i<0$, which makes the master equation not of the Lindblad type. 
The algorithm works by introducing a signed bit $s(t)$ which takes on the values $\pm 1$ and evolves with time. 
The bit is initialized at $s(t_0) = +1$. Following standard unraveling schemes, the algorithm evolves pure states with an effective Hamiltonian $H_\text{eff}$ similar to Eq.~\eqref{eq:Heff}. 
For each jump operator $L_i$, a jump probability is defined, which is proportional to $\braket{\psi(t)|L^\dagger_i L_i|\psi(t)}$. 
Based on the jump probabilities, at each timestep, a random draw determines whether a jump is applied or the state is evolved with the effective Hamiltonian. 
If a jump with $\gamma_i<0$ is applied, the signed bit is flipped $s(t+\delta t) = \gamma_i/|\gamma_i|s(t)$. 
The final solution is then obtained from the weighted average 
\begin{equation}
    \rho(t) = \mathbb{E}\left[s(t)\ketbra{\psi(t)}{\psi(t)}\right],
    \label{eq:plqt_avg}
\end{equation}
while expectation values of observables can be calculated as 
\begin{equation}
   \braket{O} = \mathbb{E}\left[s(t)\braket{\psi(t)|O|\psi(t)}\right],
   \label{eq:plqt_avg_obs}
\end{equation}
where averages are understood to be over independent pure state trajectories $\ket{\psi(t)}$.
For more details, we refer the reader to Ref.~\cite{Becker_2023}.

The algorithm in Ref.~\cite{Becker_2023} uses a simple expansion of the time evolution operator $\exp(-\mathrm{i}\delta t H_\text{eff})$ at first order in the time step $\delta t$.  
In principle, we can extend the algorithm beyond first order in the time discretization by using a waiting-time approach~\cite{Daley_2014}. 
In the waiting-time approach, a single trajectory is evolved with $H_\text{eff}$ until the state norm falls below a given threshold, triggering a quantum jump. 
The evolution up to the jump trigger can then be carried out with any solver, including implicit schemes such as Crank--Nicolson, as used in the previous implementation of the QTraj code. 
Such implicit solvers are especially useful for the evolution with non-Hermitian Hamiltonians.
To extend the PLQT algorithm to the waiting-time approach, it is necessary to introduce an additional survival-probability variable, denoted by $R(t)$. 
The need for this stems from the fact that, in the ordinary waiting-time approach, the state's norm is used to track the jump probability. 
In the PLQT case, this is not possible, as negative components of the width can increase the norm of the state during the $H_\text{eff}$ evolution. 
Thus, the state's norm no longer matches the jump probability. 
We bypass this by checking the jump criterion based on the survival probability $R(t)$, which decays with the sum of all jump probabilities at each time step. 

While this enables us to apply the PLQT algorithm at the same order of $\delta t$ as the Lindblad unravelings in QTraj, it introduces an additional potential computational cost. An average of the type~\eqref{eq:plqt_avg} is prone to a sign problem, if $\mathbb{E}[s(t)]\to 0$. 
As shown in the appendix of Ref.~\cite{Becker_2023}, $\mathbb{E}[s(t)]$ is expected to decay exponentially, with an effective sign-decay timescale $\tau_s$ defined by
\begin{equation}
    \frac{t}{\tau_s} \propto \int^t_0 \dd t^\prime\sum_{i,\gamma_i<0} \braket{\psi(t^\prime)|L^\dagger_iL_i|\psi(t^\prime)},
\end{equation}
with the sum only running over operators with negative sign in the master equation. 
The timescale $\tau_s$ defines the range in which the time evolution can be simulated efficiently with the PLQT algorithm, and it becomes shorter if contributions from negative jump operators are larger. 
This observation motivated the original introduction of optimal operator rotations~\cite{Becker_2023,Becker:2023qvq}, which directly extend the timescale.
In the case of the bottomonium master equation, the evolution of $\mathbb{E}[s(t)]$ depends on the color and angular momentum sector. 
Considering only trajectories in a singlet state with $l=0$, as only these contribute to the S-wave overlaps, we find that $\mathbb{E}[s(t)]$ only decays slowly, while for higher angular momenta and trajectories in an octet state, $\mathbb{E}[s(t)]$ decays faster. 
In principle, for the trajectories that contribute to the S-wave overlaps, $\tau_s$ should therefore be long enough for practical simulations. 

However, there is a second source of additional cost. 
As described in Ref.~\cite{Becker_2023}, the norm of a trajectory can increase. 
This means that, while the signed ensemble average preserves $\text{Tr}\rho(t)=1$ in the infinite-trajectory limit, individual trajectories can have different norms and therefore contribute with different magnitudes to the averages~\eqref{eq:plqt_avg} and~\eqref{eq:plqt_avg_obs}. 
In general, the increase of the norm is dictated by the same negative-sector activity $\braket{\psi(t)|L^\dagger_i L_i|\psi(t)}$ that controls the decay timescale~$\tau_s$. As mentioned previously, this factor can become large for high angular momenta. 
Hence, trajectories that jump to high angular momenta and later contribute to the S-wave overlaps by returning to $l=0$ can obtain a large norm. 
While such trajectories are rare, due to their large norm, they give disproportionately large contributions to the overlap~\eqref{eq:plqt_avg_obs}. 
The distribution of overlaps, which is sampled to obtain the expectation $\braket{O}$, therefore obtains a heavy tail. 
Heavy-tailed distributions are prone to a very large variance. 
Although the variance is finite because the norm growth is bounded, it can become very large in practice. 
While simulations at larger lattice spacings are possible, this problem is especially severe for small lattice spacings, as $\braket{\psi(t)|L^\dagger_i L_i|\psi(t)}$ grows with decreasing lattice spacing, making large-scale simulations impractical. 
We note that the simulation with $H_\text{eff}$ alone is not affected by these issues, as no stochastic averaging is required. 
Since the 1S survival probabilities generally do not receive notable corrections from jump processes, we can efficiently obtain results for them, including the negative terms, in the simulation. 

\end{appendix}

\bibliographystyle{apsrev4-1}
\renewcommand*{\bibfont}{\footnotesize}
\bibliography{biblio.bib}

@article{Brambilla:2016wgg,
    author = "Brambilla, Nora and Escobedo, Miguel A. and Soto, Joan and Vairo, Antonio",
    title = "{Quarkonium suppression in heavy-ion collisions: an open quantum system approach}",
    eprint = "1612.07248",
    archivePrefix = "arXiv",
    primaryClass = "hep-ph",
    reportNumber = "ICCUB-16-044, TUM-EFT-55-14",
    doi = "10.1103/PhysRevD.96.034021",
    journal = "Phys. Rev. D",
    volume = "96",
    number = "3",
    pages = "034021",
    year = "2017"
}

@article{Brambilla:2017zei,
    author = "Brambilla, Nora and Escobedo, Miguel A. and Soto, Joan and Vairo, Antonio",
    title = "{Heavy quarkonium suppression in a fireball}",
    eprint = "1711.04515",
    archivePrefix = "arXiv",
    primaryClass = "hep-ph",
    reportNumber = "TUM-EFT-89-16",
    doi = "10.1103/PhysRevD.97.074009",
    journal = "Phys. Rev. D",
    volume = "97",
    number = "7",
    pages = "074009",
    year = "2018"
}

@article{Brambilla:2022ynh,
    author = "Brambilla, Nora and Escobedo, Miguel {\'A}ngel and Islam, Ajaharul and Strickland, Michael and Tiwari, Anurag and Vairo, Antonio and Vander Griend, Peter",
    title = "{Heavy quarkonium dynamics at next-to-leading order in the binding energy over temperature}",
    eprint = "2205.10289",
    archivePrefix = "arXiv",
    primaryClass = "hep-ph",
    reportNumber = "TUM-EFT 169/22",
    doi = "10.1007/JHEP08(2022)303",
    journal = "JHEP",
    volume = "08",
    pages = "303",
    year = "2022"
}

@article{Daley_2014,
   title={Quantum trajectories and open many-body quantum systems},
   volume={63},
   ISSN={1460-6976},
   url={http://dx.doi.org/10.1080/00018732.2014.933502},
   DOI={10.1080/00018732.2014.933502},
   number={2},
   journal={Advances in Physics},
   publisher={Informa UK Limited},
   author={Daley, Andrew J.},
   year={2014},
   month=mar, pages={77–149} }

@article{Becker:2021xcj,
    author = "Becker, Tobias and Wu, Ling-Na and Eckardt, Andr{\'e}",
    title = "{Lindbladian approximation beyond ultra-weak coupling}",
    eprint = "2012.14208",
    archivePrefix = "arXiv",
    primaryClass = "quant-ph",
    doi = "10.1103/PhysRevE.104.014110",
    journal = "Phys. Rev. E",
    volume = "104",
    pages = "014110",
    year = "2021"
}

@article{Becker_2023,
    author = "Becker, Tobias and Netzer, Ch{\'e} and Eckardt, Andr{\'e}",
    title = "{Quantum Trajectories for Time-Local Non-Lindblad Master Equations}",
    eprint = "2306.14876",
    archivePrefix = "arXiv",
    primaryClass = "quant-ph",
    doi = "10.1103/PhysRevLett.131.160401",
    journal = "Phys. Rev. Lett.",
    volume = "131",
    number = "16",
    pages = "160401",
    year = "2023"
}

@article{Becker:2023qvq,
    author = "Becker, Tobias and Eckardt, Andr{\'e}",
    title = "{Optimal form of time-local non-Lindblad master equations}",
    eprint = "2312.15066",
    archivePrefix = "arXiv",
    primaryClass = "quant-ph",
    doi = "10.1103/PhysRevE.111.034132",
    journal = "Phys. Rev. E",
    volume = "111",
    number = "3",
    pages = "034132",
    year = "2025"
}

@article{Lindblad:1975ef,
    author = "Lindblad, Goran",
    title = "{On the Generators of Quantum Dynamical Semigroups}",
    reportNumber = "TRITA-TFY-75-1",
    doi = "10.1007/BF01608499",
    journal = "Commun. Math. Phys.",
    volume = "48",
    pages = "119",
    year = "1976"
}

@article{Gorini:1975nb,
    author = "Gorini, Vittorio and Kossakowski, Andrzej and Sudarshan, E. C. G.",
    title = "{Completely Positive Dynamical Semigroups of N Level Systems}",
    reportNumber = "CPT-244-TEXAS, ORO-3992-200",
    doi = "10.1063/1.522979",
    journal = "J. Math. Phys.",
    volume = "17",
    pages = "821",
    year = "1976"
}

@article{Omar:2021kra,
    author = "Omar, Hisham Ba and Escobedo, Miguel {\'A}ngel and Islam, Ajaharul and Strickland, Michael and Thapa, Sabin and Vander Griend, Peter and Weber, Johannes Heinrich",
    title = "{QTRAJ 1.0: A Lindblad equation solver for heavy-quarkonium dynamics}",
    eprint = "2107.06147",
    archivePrefix = "arXiv",
    primaryClass = "physics.comp-ph",
    reportNumber = "TUM-EFT 142/21; HU-EP-21/17-RTG",
    doi = "10.1016/j.cpc.2021.108266",
    journal = "Comput. Phys. Commun.",
    volume = "273",
    pages = "108266",
    year = "2022"
}

@article{Breuer:1999ma,
    author = "Breuer, H. P. and Kappler, B. and Petruccione, F.",
    title = "{Stochastic wave function method for non-markovian quantum master equations}",
    eprint = "quant-ph/9906024",
    archivePrefix = "arXiv",
    doi = "10.1103/PhysRevA.59.1633",
    journal = "Phys. Rev. A",
    volume = "59",
    pages = "1633--1643",
    year = "1999"
}

@article{Breuer:2004cdm,
    author = "Breuer, Heinz-Peter",
    title = "{Genuine quantum trajectories for non-Markovian processes}",
    eprint = "quant-ph/0403117",
    archivePrefix = "arXiv",
    doi = "10.1103/PhysRevA.70.012106",
    journal = "Phys. Rev. A",
    volume = "70",
    pages = "012106",
    year = "2004"
}

@article{Donvil:2021bwe,
    author = "Donvil, Brecht and Muratore-Ginanneschi, Paolo",
    title = "{Quantum trajectory framework for general time-local master equations}",
    eprint = "2102.10355",
    archivePrefix = "arXiv",
    primaryClass = "quant-ph",
    doi = "10.1038/s41467-022-31533-8",
    journal = "Nature Commun.",
    volume = "13",
    number = "1",
    pages = "4140",
    year = "2022"
}

@article{Kondov_2003,
   title={Stochastic unraveling of Redfield master equations and its application to electron transfer problems},
   volume={119},
   ISSN={1089-7690},
   url={http://dx.doi.org/10.1063/1.1605095},
   DOI={10.1063/1.1605095},
   number={13},
   journal={The Journal of Chemical Physics},
   publisher={AIP Publishing},
   author={Kondov, Ivan and Kleinekathöfer, Ulrich and Schreiber, Michael},
   year={2003},
   month=oct, pages={6635–6646} }

@article{HotQCD:2025fbd,
    author = "Bollweg, Dennis and Dasilva Gol{\'a}n, Jorge Luis and Kaczmarek, Olaf and Larsen, Rasmus Norman and Moore, Guy D. and Mukherjee, Swagato and Petreczky, Peter and Shu, Hai-Tao and Stendebach, Simon and Weber, Johannes Heinrich",
    collaboration = "HotQCD",
    title = "{Temperature dependence of heavy quark diffusion from (2+1)-flavor lattice QCD}",
    eprint = "2506.11958",
    archivePrefix = "arXiv",
    primaryClass = "hep-lat",
    doi = "10.1007/JHEP09(2025)180",
    journal = "JHEP",
    volume = "09",
    pages = "180",
    year = "2025"
}

@article{Brambilla:2019tpt,
    author = "Brambilla, Nora and Escobedo, Miguel A. and Vairo, Antonio and Vander Griend, Peter",
    title = "{Transport coefficients from in medium quarkonium dynamics}",
    eprint = "1903.08063",
    archivePrefix = "arXiv",
    primaryClass = "hep-ph",
    reportNumber = "TUM-EFT 122/18",
    doi = "10.1103/PhysRevD.100.054025",
    journal = "Phys. Rev. D",
    volume = "100",
    number = "5",
    pages = "054025",
    year = "2019"
}

@article{Ali:2025iux,
    author = "Ali, Sajid and Bala, Dibyendu and Kaczmarek, Olaf and Pavan",
    collaboration = "HotQCD",
    title = "{Thermal static potential and pseudoscalar quarkonium spectral functions from (2+1)-flavor lattice QCD}",
    eprint = "2505.11313",
    archivePrefix = "arXiv",
    primaryClass = "hep-lat",
    doi = "10.1103/s8gw-n43f",
    journal = "Phys. Rev. D",
    volume = "112",
    number = "5",
    pages = "054510",
    year = "2025"
}

@article{Ding:2025fvo,
    author = "Ding, Heng-Tong and Huang, Wei-Ping and Larsen, Rasmus and Meinel, Stefan and Mukherjee, Swagato and Petreczky, Peter and Tang, Zhanduo",
    title = "{In-medium bottomonium properties from lattice NRQCD calculations with extended meson operators}",
    eprint = "2501.11257",
    archivePrefix = "arXiv",
    primaryClass = "hep-lat",
    doi = "10.1007/JHEP05(2025)149",
    journal = "JHEP",
    volume = "05",
    pages = "149",
    year = "2025"
}

@article{BR:2025lhx,
    author = "B. R., Vyshakh and Sharma, Rishi",
    title = "{Non-Markovian dynamics of bottomonia in the QGP}",
    eprint = "2504.19348",
    archivePrefix = "arXiv",
    primaryClass = "hep-ph",
    doi = "10.1103/1tjb-hkwj",
    journal = "Phys. Rev. D",
    volume = "112",
    number = "11",
    pages = "114038",
    year = "2025"
}

@article{Brambilla:2023hkw,
    author = "Brambilla, Nora and Escobedo, Miguel {\'A}ngel and Islam, Ajaharul and Strickland, Michael and Tiwari, Anurag and Vairo, Antonio and Vander Griend, Peter",
    title = "{Regeneration of bottomonia in an open quantum systems approach}",
    eprint = "2302.11826",
    archivePrefix = "arXiv",
    primaryClass = "hep-ph",
    reportNumber = "TUM-EFT 178/23, FERMILAB-PUB-23-060-V, TUM-EFT 178/23; FERMILAB-PUB-23-060-V",
    doi = "10.1103/PhysRevD.108.L011502",
    journal = "Phys. Rev. D",
    volume = "108",
    number = "1",
    pages = "L011502",
    year = "2023"
}

@article{Brambilla:2024tqg,
    author = "Brambilla, Nora and Magorsch, Tom and Strickland, Michael and Vairo, Antonio and Vander Griend, Peter",
    title = "{Bottomonium suppression from the three-loop QCD potential}",
    eprint = "2403.15545",
    archivePrefix = "arXiv",
    primaryClass = "hep-ph",
    reportNumber = "TUM-EFT 186/23, FERMILAB-PUB-24-0029-V",
    doi = "10.1103/PhysRevD.109.114016",
    journal = "Phys. Rev. D",
    volume = "109",
    number = "11",
    pages = "114016",
    year = "2024"
}

@article{Matsui:1986dk,
    author = "Matsui, T. and Satz, H.",
    title = "{$J/\psi$ Suppression by Quark-Gluon Plasma Formation}",
    reportNumber = "BNL-38344",
    doi = "10.1016/0370-2693(86)91404-8",
    journal = "Phys. Lett. B",
    volume = "178",
    pages = "416--422",
    year = "1986"
}

@article{Datta:2025tpf,
    author = "Datta, Saumen and Banerjee, Debasish and Brambilla, Nora and Janer, Marc and Leino, Viljami and Mayer-Steudte, Julian and Petreczky, Peter and Singh, Balbeer and Vairo, Antonio",
    title = "{Lattice study of correlators for quarkonium decay}",
    eprint = "2506.22594",
    archivePrefix = "arXiv",
    primaryClass = "hep-lat",
    reportNumber = "TIFR/TH/25-13",
    doi = "10.1016/j.jspc.2025.100156",
    journal = "J. Subatomic Part. Cosmol.",
    volume = "4",
    pages = "100156",
    year = "2025"
}

@article{Brambilla:2025cqy,
    author = "Brambilla, Nora and Datta, Saumen and Janer, Marc and Leino, Viljami and Mayer-Steudte, Julian and Petreczky, Peter and Vairo, Antonio",
    collaboration = "TUMQCD",
    title = "{Lattice study of correlators of chromoelectric fields for heavy quarkonium dynamics in the quark-gluon plasma}",
    eprint = "2505.16603",
    archivePrefix = "arXiv",
    primaryClass = "hep-lat",
    reportNumber = "TUM-EFT 189/24, MITP-24-077",
    doi = "10.1103/387k-mdl1",
    journal = "Phys. Rev. D",
    volume = "112",
    number = "7",
    pages = "074509",
    year = "2025"
}

@article{Andronic:2024oxz,
    author = "Andronic, A. and others",
    title = "{Comparative study of quarkonium transport in hot QCD matter}",
    eprint = "2402.04366",
    archivePrefix = "arXiv",
    primaryClass = "nucl-th",
    reportNumber = "FERMILAB-PUB-24-0005-T-V",
    doi = "10.1140/epja/s10050-024-01306-6",
    journal = "Eur. Phys. J. A",
    volume = "60",
    number = "4",
    pages = "88",
    year = "2024"
}

@article{Brambilla:2004jw,
    author = "Brambilla, Nora and Pineda, Antonio and Soto, Joan and Vairo, Antonio",
    title = "{Effective Field Theories for Heavy Quarkonium}",
    eprint = "hep-ph/0410047",
    archivePrefix = "arXiv",
    reportNumber = "IFUM-805-FT, UB-ECM-PF-04-24",
    doi = "10.1103/RevModPhys.77.1423",
    journal = "Rev. Mod. Phys.",
    volume = "77",
    pages = "1423",
    year = "2005"
}

@book{Breuer:2002pc,
    author = "Breuer, H. P. and Petruccione, F.",
    title = "{The theory of open quantum systems}",
    year = "2002"
}

@article{Akamatsu:2020ypb,
    author = "Akamatsu, Yukinao",
    title = "{Quarkonium in quark{\textendash}gluon plasma: Open quantum system approaches re-examined}",
    eprint = "2009.10559",
    archivePrefix = "arXiv",
    primaryClass = "nucl-th",
    doi = "10.1016/j.ppnp.2021.103932",
    journal = "Prog. Part. Nucl. Phys.",
    volume = "123",
    pages = "103932",
    year = "2022"
}

@article{Yao:2021lus,
    author = "Yao, Xiaojun",
    title = "{Open quantum systems for quarkonia}",
    eprint = "2102.01736",
    archivePrefix = "arXiv",
    primaryClass = "hep-ph",
    reportNumber = "MIT-CTP/5273",
    doi = "10.1142/S0217751X21300106",
    journal = "Int. J. Mod. Phys. A",
    volume = "36",
    number = "20",
    pages = "2130010",
    year = "2021"
}

@article{Blaizot:2017ypk,
    author = "Blaizot, Jean-Paul and Escobedo, Miguel Angel",
    title = "{Quantum and classical dynamics of heavy quarks in a quark-gluon plasma}",
    eprint = "1711.10812",
    archivePrefix = "arXiv",
    primaryClass = "hep-ph",
    doi = "10.1007/JHEP06(2018)034",
    journal = "JHEP",
    volume = "06",
    pages = "034",
    year = "2018"
}

@article{Sharma:2019xum,
    author = "Sharma, Rishi and Tiwari, Anurag",
    title = "{Quantum evolution of quarkonia with correlated and uncorrelated noise}",
    eprint = "1912.07036",
    archivePrefix = "arXiv",
    primaryClass = "hep-ph",
    doi = "10.1103/PhysRevD.101.074004",
    journal = "Phys. Rev. D",
    volume = "101",
    number = "7",
    pages = "074004",
    year = "2020"
}

@article{Yao:2018nmy,
    author = "Yao, Xiaojun and Mehen, Thomas",
    title = "{Quarkonium in-medium transport equation derived from first principles}",
    eprint = "1811.07027",
    archivePrefix = "arXiv",
    primaryClass = "hep-ph",
    doi = "10.1103/PhysRevD.99.096028",
    journal = "Phys. Rev. D",
    volume = "99",
    number = "9",
    pages = "096028",
    year = "2019"
}

@article{Miura:2019ssi,
    author = "Miura, Takahiro and Akamatsu, Yukinao and Asakawa, Masayuki and Rothkopf, Alexander",
    title = "{Quantum Brownian motion of a heavy quark pair in the quark-gluon plasma}",
    eprint = "1908.06293",
    archivePrefix = "arXiv",
    primaryClass = "nucl-th",
    doi = "10.1103/PhysRevD.101.034011",
    journal = "Phys. Rev. D",
    volume = "101",
    number = "3",
    pages = "034011",
    year = "2020"
}

@article{Akamatsu:2014qsa,
    author = "Akamatsu, Yukinao",
    title = "{Heavy quark master equations in the Lindblad form at high temperatures}",
    eprint = "1403.5783",
    archivePrefix = "arXiv",
    primaryClass = "hep-ph",
    doi = "10.1103/PhysRevD.91.056002",
    journal = "Phys. Rev. D",
    volume = "91",
    number = "5",
    pages = "056002",
    year = "2015"
}

@article{PhysRevLett.68.580,
  title = {Wave-function approach to dissipative processes in quantum optics},
  author = {Dalibard, Jean and Castin, Yvan and M\o{}lmer, Klaus},
  journal = {Phys. Rev. Lett.},
  volume = {68},
  issue = {5},
  pages = {580--583},
  numpages = {0},
  year = {1992},
  month = {Feb},
  publisher = {American Physical Society},
  doi = {10.1103/PhysRevLett.68.580},
  url = {https://link.aps.org/doi/10.1103/PhysRevLett.68.580}
}

@article{ALICE:2020wwx,
    author = "Acharya, Shreyasi and others",
    collaboration = "ALICE",
    title = "{{\Upsilon} production and nuclear modification at forward rapidity in Pb{\textendash}Pb collisions at sNN=5.02TeV}",
    eprint = "2011.05758",
    archivePrefix = "arXiv",
    primaryClass = "nucl-ex",
    reportNumber = "CERN-EP-2020-207",
    doi = "10.1016/j.physletb.2021.136579",
    journal = "Phys. Lett. B",
    volume = "822",
    pages = "136579",
    year = "2021"
}

@article{ATLAS:2022exb,
    author = "Aad, Georges and others",
    collaboration = "ATLAS",
    title = "{Production of {\ensuremath{\Upsilon}}(nS) mesons in Pb+Pb and pp collisions at 5.02 TeV}",
    eprint = "2205.03042",
    archivePrefix = "arXiv",
    primaryClass = "nucl-ex",
    reportNumber = "CERN-EP-2022-045",
    doi = "10.1103/PhysRevC.107.054912",
    journal = "Phys. Rev. C",
    volume = "107",
    number = "5",
    pages = "054912",
    year = "2023"
}

@article{CMS:2018zza,
    author = "Sirunyan, Albert M and others",
    collaboration = "CMS",
    title = "{Measurement of nuclear modification factors of $\Upsilon$(1S), $\Upsilon$(2S), and $\Upsilon$(3S) mesons in PbPb collisions at $\sqrt{s_{_\mathrm{NN}}} =$ 5.02 TeV}",
    eprint = "1805.09215",
    archivePrefix = "arXiv",
    primaryClass = "hep-ex",
    reportNumber = "CMS-HIN-16-023, CERN-EP-2018-110",
    doi = "10.1016/j.physletb.2019.01.006",
    journal = "Phys. Lett. B",
    volume = "790",
    pages = "270--293",
    year = "2019"
}

@article{CMS:2023lfu,
    author = "Tumasyan, Armen and others",
    collaboration = "CMS",
    title = "{Observation of the {\Upsilon}(3S) Meson and Suppression of {\Upsilon} States in Pb-Pb Collisions at sNN=5.02{\,}{\,}TeV}",
    eprint = "2303.17026",
    archivePrefix = "arXiv",
    primaryClass = "hep-ex",
    reportNumber = "CMS-HIN-21-007, CERN-EP-2023-011",
    doi = "10.1103/PhysRevLett.133.022302",
    journal = "Phys. Rev. Lett.",
    volume = "133",
    number = "2",
    pages = "022302",
    year = "2024"
}

@article{STAR:2022rpk,
    author = "Aboona, Bassam and others",
    collaboration = "STAR",
    title = "{Observation of sequential $\Upsilon$ suppression in Au+Au collisions at $\sqrt{s_{_\mathrm{NN}}}$ = 200 GeV with the STAR experiment}",
    eprint = "2207.06568",
    archivePrefix = "arXiv",
    primaryClass = "nucl-ex",
    doi = "10.1103/PhysRevLett.130.112301",
    journal = "Phys. Rev. Lett.",
    volume = "130",
    number = "11",
    pages = "112301",
    year = "2023"
}

@inproceedings{Skullerud:2026sek,
    author = "Skullerud, Jon-Ivar and others",
    title = "{Heavy quark thermodynamics with anisotropic lattices}",
    booktitle = "{Excited QCD 2026 Workshop}",
    eprint = "2604.20449",
    archivePrefix = "arXiv",
    primaryClass = "hep-lat",
    month = "4",
    year = "2026"
}

@article{Tang:2024dkz,
    author = "Tang, Zhanduo and Mukherjee, Swagato and Petreczky, Peter and Rapp, Ralf",
    title = "{Bottomonium properties in the quark-gluon plasma from a lattice-QCD informed T-matrix approach}",
    eprint = "2411.09132",
    archivePrefix = "arXiv",
    primaryClass = "nucl-th",
    doi = "10.1103/tdc8-mf37",
    journal = "Phys. Rev. D",
    volume = "112",
    number = "3",
    pages = "034030",
    year = "2025"
}

@article{Boguslavski:2018beu,
    author = "Boguslavski, K. and Kurkela, A. and Lappi, T. and Peuron, J.",
    title = "{Spectral function for overoccupied gluodynamics from real-time lattice simulations}",
    eprint = "1804.01966",
    archivePrefix = "arXiv",
    primaryClass = "hep-ph",
    reportNumber = "CERN-TH-2018-075",
    doi = "10.1103/PhysRevD.98.014006",
    journal = "Phys. Rev. D",
    volume = "98",
    number = "1",
    pages = "014006",
    year = "2018"
}

@article{Boguslavski:2022dee,
    author = {Boguslavski, Kirill and Hotzy, Paul and M{\"u}ller, David I.},
    title = "{Stabilizing complex Langevin for real-time gauge theories with an anisotropic kernel}",
    eprint = "2212.08602",
    archivePrefix = "arXiv",
    primaryClass = "hep-lat",
    doi = "10.1007/JHEP06(2023)011",
    journal = "JHEP",
    volume = "06",
    pages = "011",
    year = "2023"
}

@article{Banuls:2015sta,
    author = "Ba{\~n}uls, M. C. and Cichy, K. and Cirac, J. I. and Jansen, K. and Saito, H.",
    title = "{Thermal evolution of the Schwinger model with Matrix Product Operators}",
    eprint = "1505.00279",
    archivePrefix = "arXiv",
    primaryClass = "hep-lat",
    reportNumber = "DESY-15-060, SFB-CPP-14-125, DESY 15-060, SFB/CPP-14-125",
    doi = "10.1103/PhysRevD.92.034519",
    journal = "Phys. Rev. D",
    volume = "92",
    number = "3",
    pages = "034519",
    year = "2015"
}

@article{Martinez:2016yna,
    author = "Martinez, E. A. and others",
    title = "{Real-time dynamics of lattice gauge theories with a few-qubit quantum computer}",
    eprint = "1605.04570",
    archivePrefix = "arXiv",
    primaryClass = "quant-ph",
    doi = "10.1038/nature18318",
    journal = "Nature",
    volume = "534",
    pages = "516--519",
    year = "2016"
}

@article{Bauer:2022hpo,
    author = "Bauer, Christian W. and others",
    title = "{Quantum Simulation for High-Energy Physics}",
    eprint = "2204.03381",
    archivePrefix = "arXiv",
    primaryClass = "quant-ph",
    reportNumber = "UMD-PP-022-04, LA-UR-22-22100, RIKEN-iTHEMS-Report-22, RIKEN-iTHEMS-Report-22,
  FERMILAB-PUB-22-249-SQMS-T, FERMILAB-PUB-22-249-SQMS-T, IQuS@UW-21-027, MITRE-21-03848-2, FERMILAB-PUB-22-249-SQMS-T",
    doi = "10.1103/PRXQuantum.4.027001",
    journal = "PRX Quantum",
    volume = "4",
    number = "2",
    pages = "027001",
    year = "2023"
}

@article{Burnier:2010rp,
    author = "Burnier, Y. and Laine, M. and Langelage, J. and Mether, L.",
    title = "{Colour-electric spectral function at next-to-leading order}",
    eprint = "1006.0867",
    archivePrefix = "arXiv",
    primaryClass = "hep-ph",
    reportNumber = "BI-TP-2010-16",
    doi = "10.1007/JHEP08(2010)094",
    journal = "JHEP",
    volume = "08",
    pages = "094",
    year = "2010"
}

@unpublished{Brambilla:2026213, 
author = {Nora Brambilla and Panayiotis Panayiotou and Saga Säppi and Antonio Vairo},
title = {{Transport coefficients for Lindbladian quarkonium
evolution in quark–gluon plasma}}, 
note = {in preparation, TUM-EFT 213/26} 
}

@unpublished{Brambilla:2024191, 
author = {Nora Brambilla and Miguel Angel Escobedo and Antonio Vairo and Peter Vander Griend}, 
title = {{Anatomy of quarkonium transport and transport
coefficients}}, 
note = {in preparation, TUM-EFT 191/24, FERMILAB-PUB-24-0451-T} 
}

@article{Binder:2021otw,
    author = "Binder, Tobias and Mukaida, Kyohei and Scheihing-Hitschfeld, Bruno and Yao, Xiaojun",
    title = "{Non-Abelian electric field correlator at NLO for dark matter relic abundance and quarkonium transport}",
    eprint = "2107.03945",
    archivePrefix = "arXiv",
    primaryClass = "hep-ph",
    reportNumber = "KEK-TH-2334, MIT-CTP/5313",
    doi = "10.1007/JHEP01(2022)137",
    journal = "JHEP",
    volume = "01",
    pages = "137",
    year = "2022"
}

@article{Scheihing-Hitschfeld:2023tuz,
    author = "Scheihing-Hitschfeld, Bruno and Yao, Xiaojun",
    title = "{Real time quarkonium transport coefficients in open quantum systems from Euclidean QCD}",
    eprint = "2306.13127",
    archivePrefix = "arXiv",
    primaryClass = "hep-ph",
    reportNumber = "INT-PUB-23-019, MIT-CTP/5575, IQuS@UW-21-051",
    doi = "10.1103/PhysRevD.108.054024",
    journal = "Phys. Rev. D",
    volume = "108",
    number = "5",
    pages = "054024",
    year = "2023",
    note = "[Erratum: Phys.Rev.D 109, 099902 (2024)]"
}

@article{Brambilla:2025xnw,
    author = {Brambilla, Nora and Panayiotou, Panayiotis and S{\"a}ppi, Saga and Vairo, Antonio},
    title = "{The chromoelectric adjoint correlators in Euclidean space at next-to-leading order}",
    eprint = "2505.16604",
    archivePrefix = "arXiv",
    primaryClass = "hep-ph",
    reportNumber = "TUM-EFT 190/24",
    doi = "10.1007/JHEP08(2025)219",
    journal = "JHEP",
    volume = "08",
    pages = "219",
    year = "2025"
}

@article{Brambilla:1999xf,
    author = "Brambilla, Nora and Pineda, Antonio and Soto, Joan and Vairo, Antonio",
    title = "{Potential NRQCD: An Effective theory for heavy quarkonium}",
    eprint = "hep-ph/9907240",
    archivePrefix = "arXiv",
    reportNumber = "CERN-TH-99-199, HEPHY-PUB-716-99, UB-ECM-PF-99-06, UWTHPH-1999-34, UB-ECM-PF-99-13",
    doi = "10.1016/S0550-3213(99)00693-8",
    journal = "Nucl. Phys. B",
    volume = "566",
    pages = "275",
    year = "2000"
}

@article{Scheihing-Hitschfeld:2022xqx,
    author = "Scheihing-Hitschfeld, Bruno and Yao, Xiaojun",
    title = "{Gauge Invariance of Non-Abelian Field Strength Correlators: The Axial Gauge Puzzle}",
    eprint = "2205.04477",
    archivePrefix = "arXiv",
    primaryClass = "hep-ph",
    reportNumber = "MIT-CTP/5425, IQuS@UW-21-040",
    doi = "10.1103/PhysRevLett.130.052302",
    journal = "Phys. Rev. Lett.",
    volume = "130",
    number = "5",
    pages = "052302",
    year = "2023"
}

@article{Nijs:2023dbc,
    author = "Nijs, Govert and Scheihing-Hitschfeld, Bruno and Yao, Xiaojun",
    title = "{Generalized gluon distribution for quarkonium dynamics in strongly coupled N=4 Yang-Mills theory}",
    eprint = "2310.09325",
    archivePrefix = "arXiv",
    primaryClass = "hep-ph",
    reportNumber = "CERN-TH-2023-170, MIT-CTP/5629, IQuS@UW-21-065",
    doi = "10.1103/PhysRevD.109.094043",
    journal = "Phys. Rev. D",
    volume = "109",
    number = "9",
    pages = "094043",
    year = "2024"
}

@article{Hammou:2026yqh,
    author = "Hammou, Aoumeur Daddi and Blaizot, Jean-Paul and Gossiaux, Pol Bernard and Gousset, Thierry",
    title = "{General Lindblad equation for quarkonium evolution in a quark-gluon plasma}",
    eprint = "2608.15347",
    archivePrefix = "arXiv",
    primaryClass = "nucl-th",
    month = "8",
    year = "2026"
}

@article{Brambilla:2008cx,
    author = "Brambilla, Nora and Ghiglieri, Jacopo and Vairo, Antonio and Petreczky, Peter",
    title = "{Static quark-antiquark pairs at finite temperature}",
    eprint = "0804.0993",
    archivePrefix = "arXiv",
    primaryClass = "hep-ph",
    reportNumber = "BNL-NT-08-7, IFUM-916-FT",
    doi = "10.1103/PhysRevD.78.014017",
    journal = "Phys. Rev. D",
    volume = "78",
    pages = "014017",
    year = "2008"
}

\end{document}